\documentclass[twocolumn,superscriptaddress,amsmath,amssymb,aps,prb,longbibliography]{revtex4-2}
\pdfoutput=1
\usepackage{amssymb}
\usepackage{amsmath}
\usepackage{braket}
\usepackage[english]{babel}
\usepackage{graphicx,xcolor}
\usepackage{bm}

\usepackage{subfig}

\newcommand{\cred}[1]{\textcolor{black}{#1}}

\newcommand{\myeta}{\epsilon^{mg}}

\begin{document}
\title{Magnetochiral anisotropy on a quantum spin Hall edge}
\author{Youjian Chen}
\affiliation{Department of Physics, University of Virginia, Charlottesville, Virginia 22904, USA}
\author{Gary Quaresima}
\affiliation{Department of Physics, University of Virginia, Charlottesville, Virginia 22904, USA}
\author{Wenjin Zhao}
\affiliation{Kavli Institute at Cornell for Nanoscale Science, Ithaca, New York 14853, USA}
\author{Elliott Runburg}
\affiliation{Department of Physics, University of Washington, Seattle, Washington 98195, USA}
\author{David Cobden}
\affiliation{Department of Physics, University of Washington, Seattle, Washington 98195, USA}
\author{D. A. Pesin}
\affiliation{Department of Physics, University of Virginia, Charlottesville, Virginia 22904, USA}
\begin{abstract}
We develop a theory of nonlinear low-magnetic-field magnetotransport on a helical edge of a quantum spin Hall insulator due to the edge state coupling to bulk midgap states. We focus on the part of the nonlinear I-V characteristic that is odd in the applied magnetic field, and quadratic in the applied bias voltage. This part of the I-V characteristic corresponds to the resistance of the sample being dependent on the relative orientation of the current and an external magnetic field, hence represents a type of edge magnetochiral anisotropy. \cred{We identify a mechanism of the magnetochiral anisotropy related to the Hubbard interaction on the midgap state, which leads to the dependence of the scattering characteristics on the current flowing on the edge. This in turn leads to bias-voltage-dependent resistance, or equivalently conductance, hence a nonlinear I-V.} We compare the developed theory to the experiments on monolayer $\mathrm{WTe}_{2}$, and find good agreement with the developed theory.
\end{abstract}
\maketitle
\textit{Introduction --- } The edge of a quantum spin Hall insulator~\cite{kane2005quantum,bernevig2006quantum,konig2007quantum,hasan2010colloquium,qi2011topological,maciejko2011review} is a disordered interacting helical liquid. Due to the edge spin-momentum locking, its transport properties are very sensitive to the various perturbations acting on the electronic spin. In particular, they are sensitive to the external magnetic field. For voltages that are not too high, the current-voltage relation on the edge can be written as
\begin{equation}\label{eq:current}
I(V,\bm B)=G(\bm B)V+\gamma(\bm B)V^2, 
\end{equation}
where $I$ is the current flowing between the source and drain, $V$ is the applied bias voltage, $G(\bm B)$ and $\gamma(\bm B)$ are the linear and nonlinear conductances, dependent on the external magnetic field $\bm B$. 

The linear magnetoconductance, $G(\bm B)$, has been a subject of many theoretical~\cite{maciejko2010magnetoconductance,buttiker2012localization,pikulin2014,essert2015magnetotransport,chen2023linearG} and experimental ~\cite{konig2008quantum,gusev2011transport,gusev2015aharonov,molenkamp2019inplane,zhao2021determination,hamilton2021spin} studies. Often a singular, cusp-like dependence on the magnetic field at low fields is reported experimentally. In Ref.~\cite{chen2023linearG} by the present authors it was shown that at least in the case of monolayer $\mathrm{WTe}_2$ the low-$B$-field dependence of the linear conductance is adequately described with a model based on edge states hybridized with midgap localized states. 

The nonlinear conductance on a topological edge, $\gamma(\bm B)$ in Eq.~\eqref{eq:current}, has received much less attention. In the experimental work of Ref.~\cite{zhao2021determination} it was shown that, just like the linear conductance, it is a sensitive probe of the edge spin-momentum locking.

From the theoretical point of view, $\gamma(\bm B)$ is interesting because it describes the DC analog of optical rectification and frequency-doubling. As a nonlinear effect, its form is not constrained by the Onsager relations~\cite{MelroseBook}. In particular, $\gamma(\bm B)$ admits a part odd in the magnetic field, $\gamma_a(\bm B)\equiv(\gamma(\bm B)-\gamma(-\bm B))/2\neq 0$. It is known~\cite{spivak2006oddB} that $\gamma_a(\bm B)$ vanishes for strictly coherent transport of non-interacting electrons, so it can serve as a probe of interactions and dephasing in the system. Furthermore, it is the odd part of the nonlinear conductance that was found to be sensitive to the edge spin texture in Ref.~\cite{zhao2021determination}. It is thus the main focus of this work.

Broadly speaking, the nonlinear magnetoconductance describes changes of the sample's resistance proportional to the current flowing in it, depending on the relative orientation of the external magnetic field and the current flow. It thus falls under the umbrella of the magnetochiral anisotropy (MCA) effects~\cite{rikken2021mca}. This is how we are going to refer to it in what follows. 

In Ref.~\cite{zhao2021determination}, some of the present authors made an attempt to phenomenologically explain the MCA on a spin Hall edge using the current-induced spin polarization, and the associated exchange field due to electron-electron interaction~\cite{rudner2019currentgap}. Such exchange field, proportional to the edge current, acts on the edge electrons just like an additional Zeeman field, and modifies electron scattering and transport on the edge, if the relevant scattering centers are in fact located on the edge, see Ref.~\cite{zhao2021determination} for details. This simple point of view was able to very closely reproduce all the features of the MCA at large enough magnetic fields, $B\gtrsim 1\mathrm{T}$, but completely failed for smaller $B$-fields, $B\lesssim 0.5\mathrm{T}$. 

In Ref.~\cite{chen2023linearG} we proposed a model of a helical edge state hybridized with nearby midgap states localized in the bulk of the system to describe low-field linear magnetotransport. The model successfully described singular cusp-like linear mangetoconductance. Within the model, it originated from strong back-scattering induced by spin precession due to the Zeeman field on midgap states resonantly coupled to the edge states. 

In this work we generalize the model used in Ref.~\cite{chen2023linearG} to include a Hubbard interaction term in the midgap state Hamiltonian, and show that it provides a mechanism for the appearance of the edge MCA, which appears to explain the experimental observations in a monolayer $\mathrm{WTe}_2$~\cite{zhao2021determination}. \cred{As in Ref.~\cite{chen2023linearG}, we consider the experimental conditions of Ref.~\cite{zhao2021determination}. In particular, we assume that a transport measurement is performed on a flake whose linear size is large compared with the ``along-the-edge" distance between the source and drain electrodes, and that the resistance of the longer edge connecting the source and drain is much larger than the resistance of the complementary shorter edge. In the opposite ideal case of perfect symmetry between the two complementary edges a nonlinear current quadratic in the applied voltage is forbidden due to the inversion symmetry that holds for the entire sample.}

\textit{Model of a helical edge state coupled to a midgap state --- } We start with a model of a helical edge state coupled to a single Kramers-degenerate midgap state. Later we will generalize to the many-impurity case for the purpose of disorder averaging. 

The electronic Hilbert space consists of the helical edge states and two states of opposite spin associated with the midgap level.  
We will describe the right-moving spin-up states and left-moving spin-down states with annihilation operators $a_{k\sigma}$, where $k$ is the quasimomentum, and  $\sigma=\uparrow,\downarrow$. We assume that the Kramers-degenerate midgap states can be labeled with the same spin indices as the edge electrons, or annihilation operators $c_{\sigma}$~\cite{note:spin}. 

\begin{figure}[b]
    \centering
    \includegraphics[width=8.6cm]{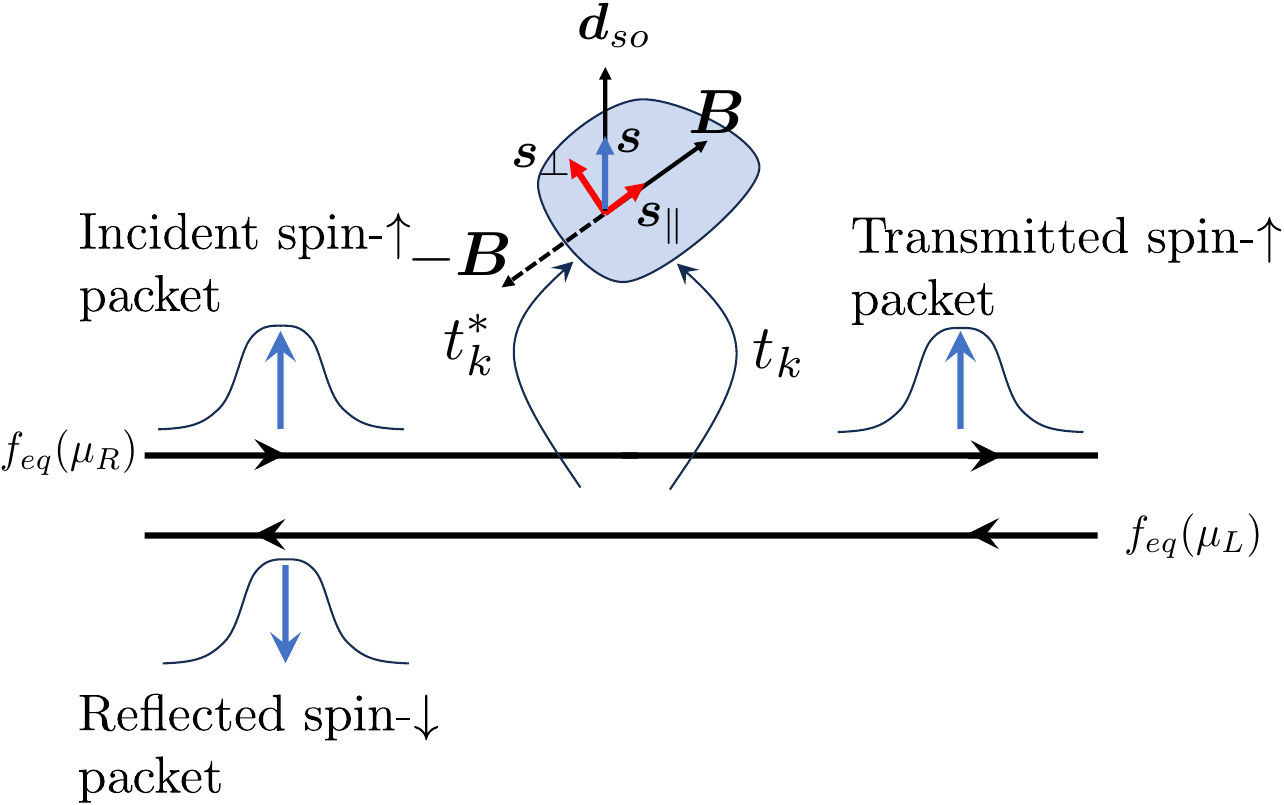}
    \caption{Schematic picture of helical electrons scattering off of a midgap state. An incident spin-$\uparrow$ wave packet tunnels onto the midgap state, precesses around the external magnetic field, then tunnels back onto the edge either as a spin-$\uparrow$ electron to continue motion in the original direction, or as a back-scattered spin-$\downarrow$ electron. }
    \label{fig:scattering}
\end{figure}

The Hamiltonian of the system is essentially given by the interacting Anderson model~\cite{anderson1961model}:
\begin{align}\label{eq:Hamiltonian}
        H=&\sum_{k \sigma} \sigma v k a^{\dagger}_{k,\sigma}a_{k,\sigma}+\sum_{k,\sigma}( t_k a^{\dagger}_{k,\sigma} c_{\sigma}+t^*_kc_{\sigma}^{\dagger} a_{k,\sigma})\nonumber \\
        &+\sum_{\sigma \sigma^{'}}\left[\myeta+(\bm b \cdot \bm{\sigma})\right]_{\sigma\sigma'}c^{\dagger}_{\sigma}c_{\sigma^{'}}+U c^{\dagger}_{\uparrow}c^{\dagger}_{\downarrow}c_{\downarrow}c_{\uparrow}.
\end{align}
The first term is the Hamiltonian for helical edge electrons with Fermi velocity $v$, in which $a_{k,\sigma}$ is the annihilation operator for the electrons with momentum $k$ and spin projection $\sigma$. The index $\sigma$ takes two values, $\sigma=\uparrow,\downarrow$, which are understood as $\pm1$, respectively, in equations. The spins of the edge electrons are directed along and opposite to $\bm d_{so}$, which defines the edge spin polarization direction, see Fig.~\ref{fig:scattering}. For the time being we will take $\bm d_{so}$ as the direction of the $z$-axis in the spin space.  The second term describes the edge-midgap state hybridization. We assume that edge-midgap state hybridization conserves spin, see note~\cite{note:spin}, and in what follows we will neglect the momentum dependence of the hybridization matrix elements, $t_k\to t$. The third term describes a midgap level with energy $\myeta$, and includes Zeeman coupling to a magnetic field $\bm {B}$. In most of the paper we will use the Zeeman field in energy units, denoted with $\bm{b}=\frac12 g_{\mathrm{mg}}\mu_{B}\bm{B}$, where $g_{\mathrm{mg}}$ is the $g$-factor for the midgap state. Again, this is a simplifying assumption about the midgap state $g$-tensor, but it seems to work as far as comparing theoretical results to experimental observation is concerned, see Ref.~\cite{chen2023linearG}, and also below. We also do not include the Zeeman field for the edge states, since its effect is small for a Fermi level away from the edge Dirac point, when there is a large spin splitting due to spin-orbit coupling.  On the midgap state the magnetic field splits two degenerate levels, hence is nonperturbative. The last term is the Hubbard interaction on the midgap level, the spin indices of operators in which are dictated by the Pauli exclusion principle. We assume that the midgap level is in the nonmagnetic limit of the Anderson model for levels relevant for scattering of edge electrons, meaning $U$ is small compared to the width of such levels, see below.

\textit{Qualitative picture of the edge MCA --- }We are interested in the odd-in-$\bm B$ nonlinear conductance, $\gamma_a(\bm B)$. Its finiteness implies that scattering of edge electrons changes when the orientation of the $\bm B$-field flips. This is strictly forbidden for linear transport by the Onsager relations.  One mechanism for the appearance of the nonlinear conductance due to the Hubbard interaction in Hamiltonian~\eqref{eq:Hamiltonian} is related to the current-induced correction to the Zeeman field on the impurity, as described below. For the purpose of a qualitative argument, we will assume small magnetic field limit, for which the spin dynamics on the impurity site is classical precession.

A scattering event for, say, an incident spin-up electron starts with a wave packet of spin-up electrons approaching the scattering region, where it then tunnels onto the midgap state. It spends the Wigner delay time~\cite{perelomov1998book} in that state.  During this time the spin of the electron precesses around the $\bm B$-field. Since the spin direction after the precession is different from the initial, the electron can tunnel back on the edge and either continue moving to the right as a spin-up electron, or back-scatter to the left as a spin-down electron. The analogous sequence of steps for the incident spin-down electrons is obvious. This backscattering mechanism explains small-field linear magnetoresistance in $\mathrm{WTe}_2$~\cite{chen2023linearG}.

Because on a current-carrying helical edge the densities of the spin-up and spin-down electrons are different (this density difference determines the current), there will be a net spin polarization on the impurity generated by the current. This polarization leads to an asymmetry of scattering under $\bm B\to -\bm B$ in the presence of electron-electron interactions. The origin of this asymmetry can be qualitatively seen from Fig.~\ref{fig:scattering} if one notices that the component of the spin polarization along the Zeeman field, which is denoted $\bm s_\parallel$ in Fig.~\ref{fig:scattering}, does not precess around it, hence does not change when the direction of the $\bm B$ flips. In the presence of an electron-electron interaction, this net spin polarization gives rise to a self-field-type correction to the external Zeeman field. This self-field due to $\bm s_\parallel$ either gets added or subtracted from the Zeeman field due to $\bm B$. For individual electrons this leads to different scattering probabilities for $\bm B$ and $-\bm B$, hence nonzero $\gamma_a(\bm B)$. Note that it is important to have a self-field created by the current, since an equilibrium one would flip its direction together with the magnetic field.  

The above argument is oversimplified, since the precise structure of the energy-resolved spin polarization on the midgap state is important, see Ref.~\cite{supplement} for details. Nevertheless, it gives an intuitive picture of the origin of the edge MCA. 

\textit{Quantitative theory of the edge MCA ---} To build a quantitative theory of the edge MCA we make the same assumptions about the edge transport as those that led to the explanation of the linear magnetotransport in monolayer $\mathrm{WTe}_2$ in our previous work~\cite{chen2023linearG}: edge transport is incoherent, each impurity is a `resistor', with the total resistance being additive; only the shortest part of the edge between the source and drain leads is taken into account, the complementary part of the edge being too resistive to carry appreciable current; the edge is long enough, and the relevant impurities are sparse enough that there is energy relaxation between scattering events, such that the distributions of the incident carriers have Fermi-Dirac form with nonequilibrium chemical potentials, different for each helical branch. These are denoted with $\mu_{L,R}$ for left- and right-movers, or, equivalently, spin-$\downarrow$ and spin-$\uparrow$ electrons in Fig.~\ref{fig:scattering}. The reader should consult Ref.~\cite{chen2023linearG} for further details. 

The above assumptions are characteristic of high-temperature transport. In what follows we always assume that the temperature is the largest energy scale in the problem, even compared to the strength of the Hubbard repulsion; see the discussion at the end of the paper. The temperature is probably not large compared to the level widths of impurities situated close to the edge, and thus strongly hybridized with it, but these do not make a substantial contribution to the linear or nonlinear magnetotransport, because transport electrons do not spend much time on them. 

We first discuss the treatment of the Hubbard repulsion in the Hartree-Fock approximation. The Hamiltonian $H_{U}=U c^{\dagger}_{\uparrow}c^{\dagger}_{\downarrow}c_{\downarrow}c_{\uparrow}$ gives rise to the following
mean-field term in the total Hamiltonian:
\begin{equation}
    H^{\rm{MF}}_{U}= U \begin{pmatrix}
     c^{\dagger}_{\uparrow} &
    c^{\dagger}_{\downarrow} \\
    \end{pmatrix} 
    \begin{pmatrix}
    \braket{c^{\dagger}_{\downarrow}c_{\downarrow}} & -\braket{c^{\dagger}_{\downarrow}c_{\uparrow}} \\
     -\braket{c^{\dagger}_{\uparrow}c_{\downarrow}} & \braket{c^{\dagger}_{\uparrow}c_{\uparrow}} \\
    \end{pmatrix}
    \begin{pmatrix}
    c_{\uparrow} \\
    c_{\downarrow}
    \end{pmatrix},
\end{equation}
where $\braket{...}$ represents an average over the nonequilibrium state of the system. The mean-field Hamiltonian contains a correction to the energy of the midgap state, $\delta\myeta$, and a correction to the Zeeman field, $\delta\bm b$: 
\begin{align}\label{eq:MFhamiltonian}
    H^{\rm{MF}}_{U}\equiv\sum_{\sigma \sigma^{'}}\left[\delta\myeta+(\delta\bm b \cdot \bm{\sigma})\right]_{\sigma\sigma'}c^{\dagger}_{\sigma}c_{\sigma^{'}},
\end{align}
where 
\begin{align}\label{eq:exchangecorrections}
    \delta \myeta=U\frac{\braket{c^{\dagger}_{\uparrow} c_{\uparrow}}+\braket{c^{\dagger}_{\downarrow} c_{\downarrow}}}{2},\quad \delta \bm b=-\frac{U}{2}\sum_{\sigma\sigma'}\bm \sigma_{\sigma\sigma'}\braket{c^{\dagger}_{\sigma} c_{\sigma'}}.
\end{align}

It is clear from the form of Hamiltonian~\eqref{eq:MFhamiltonian} that it simply modifies the midgap state energy and the Zeeman field. The averages of electronic operators that determine these corrections, appearing on the right hand side of Eqs.~\eqref{eq:exchangecorrections}, do not vanish even in equilibrium, without a transport current on the edge. In the presence of a current, they receive nonequilibrium corrections, which, to the leading order, are proportional to the current. Equivalently, they are proportional to the two-point voltage drop across an impurity, $(\mu_R-\mu_L)/|e|$, see Fig.~\ref{fig:scattering}. \cred{(For many impurities, this voltage is distinct from the applied source-drain voltage.)} The explicit form of the equilibrium and  nonequilibrium corrections to the midgap state energy and the Zeeman field is discussed extensively in Ref.~\cite{supplement}.  We will see below that a very small value of $U$ is required to fit the experimental data on nonlinear magnetotransport, so it is a good approximation to simply neglect the equilibrium interaction corrections to the midgap state Hamiltonian. \cred{The smallness of $U$ also justifies the ``single-shot" Hartree-Fock treatment, as opposite for a self-consistent one for relatively strong interaction.}

At the mean-field level, the solution to the scattering problem for a single midgap state reduces to that presented in Ref.~\cite{chen2023linearG}. We can write the expression for the reflection coefficient if we introduce the midgap level width due to coupling to the edge state,  $\Gamma(E)=L_x |t|^2/2v$. $\Gamma$ is independent of the normalization length $L_x$ since $t\propto 1/\sqrt{L_x}$. The result for the reflection coefficient can be written in terms of the total Zeeman field, $\bm b_{tot}\equiv \bm b+\delta\bm b$, and the total midgap state energy, $\myeta_{tot}\equiv \myeta+\delta \myeta$:
\begin{equation}\label{eq:reflectioncoefficient}
{\cal R}(E)=\frac{4\Gamma^2(\bm b_{tot}\times \bm z)^2}
    {((E-\myeta_{tot})^2+\Gamma^2-\bm b_{tot}^2)^2+4\Gamma^2\bm b_{tot}^2},
\end{equation}
where $E$ is the energy of the scattering electron. \cred{The numerator of this expression shows that the reflection vanishes for the magnetic field along the edge spin polarization direction. This is expected, since such a field cannot flip the spin of the edge electron. The denominator shows that two resonant midgap levels with energies $E=\myeta_{tot}\pm |\bm b_{tot}|+i\Gamma$ participate in the scattering process. }

We can obtain the nonlinear conductance, $\gamma_a(\bm B)$ in Eq.~\eqref{eq:current}, from the reflection coefficient~\eqref{eq:reflectioncoefficient} by calculating the edge resistance of many midgap states in the incoherent high-temperature regime, and averaging the result over their energies and spatial positions throughout the sample, see Section~VI of Ref.~\cite{supplement} for details. There it is argued that the correction to the midgap state energy can be neglected for small $B$-fields. 

The calculation is complicated by the fact that the value of $\delta \bm b$ depends on the impurity energy and spatial position, but is greatly simplified by the observation that the nonlinear current is in practice very small compared with the linear one~\cite{zhao2021determination}. This means that the nonlinear conductance can be obtained by simply expanding the total resistance in $\delta \bm b$ for each impurity to linear order, and averaging the result. There is no `back-action' from the nonlinear current on the linear one. 

As a result, we obtain the following small-B expression for the MCA coefficient:
\begin{align}\label{eq:gammafinal}
    \gamma_a(B,\theta)=\tilde{\gamma}B\cos\theta\sin^2\theta,
\end{align}
where $\tilde\gamma$ is a constant proportional to the length of the channel. This is the central result of this work.

We note that there is another mechanism of magnetochiral anisotropy on a topological edge, which is related to its band structure. It is shown in Section~VII of Ref.~\cite{supplement} that this mechanism effectively changes $\cos\theta\sin^2\theta$ to $\cos\theta(\sin^2\theta+\delta)$ in Eq.~\eqref{eq:gammafinal}. 
This band mechanism alone cannot explain experimental angular dependence of the nonlinear magnetoconductance.

\textit{Comparison with experiment ---} We now compare the developed theory with the measurements performed on monolayer $\mathrm{WTe}_2$~\cite{zhao2021determination}. At this point we need to take into account that the edge spin polarization in WTe$_2$ lies in the mirror plane of the material, and makes an angle with the normal to the sample, denote with $\theta_0$ below. We will also switch from the magnetic field in energy units, $b$, back to the field measured in Tesla, $B$.

The results of the entire preceding discussion can be summarized in the following expression for the small-B-field MCA on the edge, described by the odd-in-$\bm B$ part of the nonlinear conductance $\gamma(\bm B)=\gamma(B,\theta)$ in Eq.~\eqref{eq:current}:

\begin{align}\label{eq:MCAforexperiment}
    \gamma_a(B,\theta)=\tilde{\gamma}(\sin^2(\theta-\theta_0)+\delta)\cos(\theta-\theta_0)\frac{B^3}{B^2+B^2_\Gamma},
\end{align}
$\delta\neq 0$ can stem from the fluctuations of the local spin polarization axis along the edge for the MCA contribution due to coupling to midgap states, and directly from the band mechanism of MCA related to the curvature of the edge state dispersion; $B_\Gamma$ is a phenomenological magnetic field scale that corresponds to the midgap level width unrelated to the coupling to the edge, which determines the rounding of experimental data at small $B$-fields. \cred{Note that linear-in-B dependence of $\gamma_a$ in Eq.~\eqref{eq:gammafinal} is recovered if we set $B_\Gamma\to 0$. }

The best fit to the experimental data is obtained for
$\tilde\gamma\approx 1.53\times 10^{-4}\mathrm{A\,V^{-2}T^{-1}}$, $\delta\approx 0.14$, $B_\Gamma\approx 0.09\mathrm{T}$, and $\theta_0=33.33^\circ$.

\begin{figure}
    \includegraphics[width=8.6cm]{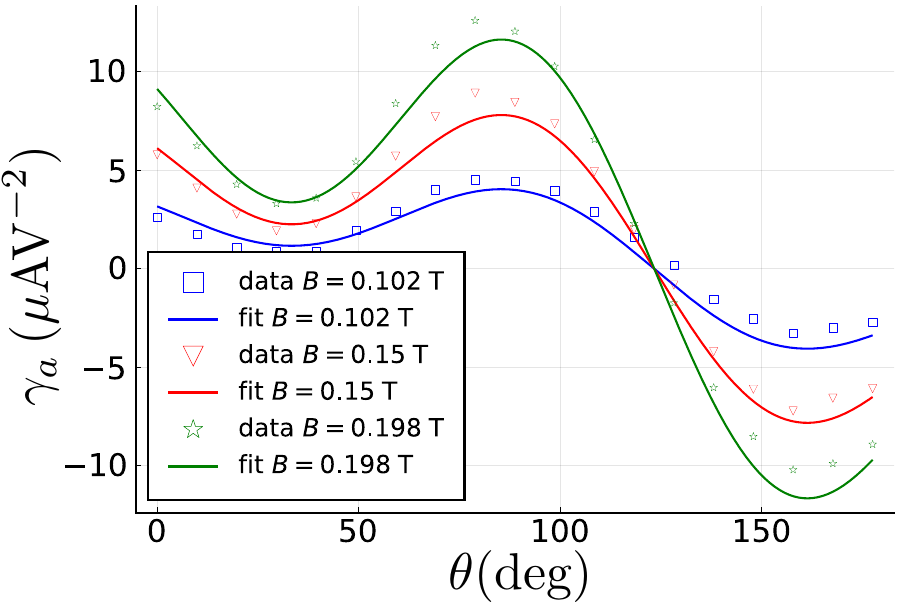}\\
    \includegraphics[width=8.6cm]{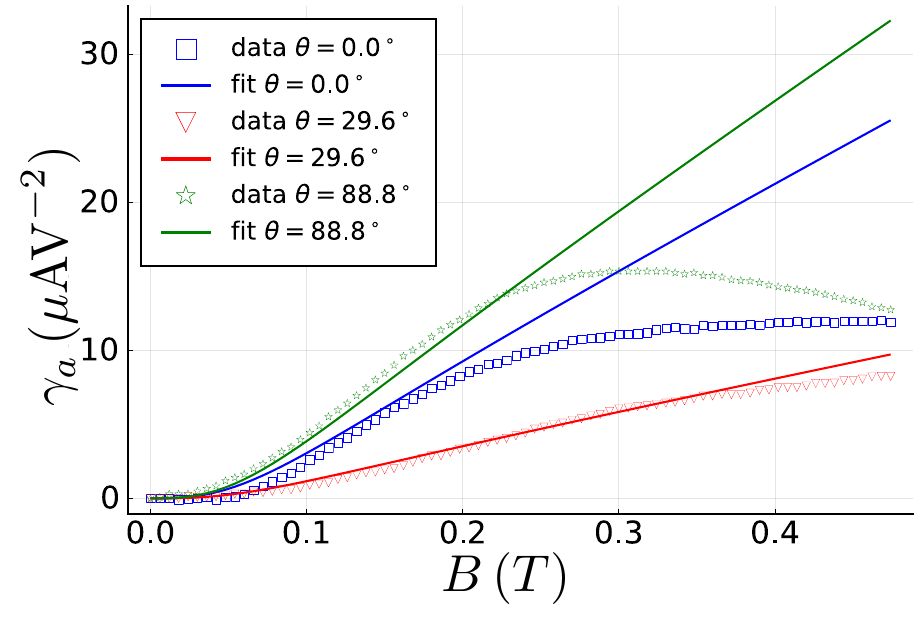}
    \caption{Experiment vs theoretical fits. Upper panel: the angular dependence of the antisymmetric part of the nonlinear conductance, $\gamma_a$, for different magnitudes of the magnetic field. Lower panel: dependence of $\gamma_a$ on the magnitude of the magnetic field for several representative angles. We note that the model works well for ``low fields" $B\lesssim 0.3\mathrm{T}$, except for when the signal is extremely small (which is when we expect the experimental data to be noisy). }
    \label{fig:MCA}
\end{figure}

The value of $\theta_0$ is very close to $33.7^\circ$ obtained Ref.~\cite{chen2023linearG}, and we checked that the same scale of the magnetic field $B_\Gamma$ determines the rounding of the linear magnetoresistance, considered, again, in Ref.~\cite{chen2023linearG}. In this sense the two new free parameters are $\tilde\gamma$ and $\delta$. A comparison between theoreticals fit and the experimental data is presented in Fig.~\ref{fig:MCA}.


\textit{Discussion --- } The main result of this work is a theoretical prediction for the angular dependence of the magnetochiral anisotropy of a quantum spin Hall edge, given by the nonlinear conductance of Eq.~\eqref{eq:MCAforexperiment}. It reproduces two experimentally observed features of this angular dependence: the nonlinear signal (almost) vanishing for the magnetic field orientation perpendicular to the edge spin polarization ($\bm B\perp \bm d_{so}$), and the nonlinear signal having local minimum for the magnetic field oriented along the edge spin polarization ($\bm B\parallel \bm d_{so}$). These properties are in fact a definition of the MCA's sensitivity to the edge spin texture. 

The first of the properties does hold for the mechanisms discussed in this paper. For the band mechanism it straightforwardly stems from the fact that the edge spectrum has a center of symmetry for $\bm B\perp \bm d_{so}$. For the mechanism based on scattering via midgap states it follows from the fact the reflection probability for that mechanism only depends on the magnitude of the angle between the total Zeeman field and $\bm d_{so}$, and the magnitude of the Zeeman field. For $\bm B\perp \bm d_{so}$, both of these quantities are even functions of $\bm B$, hence $\gamma_a$ vanishes. 

The second - a local minimum for the nonlinear current for $\bm B\parallel \bm d_{so}$ - can also be easily reproduced with the present theory. In fact, in the ideal case of strictly constant edge spin polarization along the edge, and purely linear edge spectrum, the MCA for this orientation of the magnetic field vanishes, and so does the linear magnetoconductance. As was discussed in Ref.~\cite{chen2023linearG}, fluctuations of the spin polarization along the edge lead to a finite linear magnetoresistance, and so do they for the case of the nonlinear part of the I-V characteristic. The band mechanism also leads to a finite nonlinear current in this case. Edge spin polarization fluctuations along the edge do not spoil the smallness of the nonlinear signal for $\bm B\perp \bm d_{so}$, since deviations of the local spin axis in opposite directions from the average make contributions of opposite signs to the net nonlinear resistance. 

Given that the developed theory matches the experimental data fairly well, we can use it to estimate the strength of the Hubbard repulsion, $U$, needed to fit the experimental data assuming that small-$B$ linear and nonlinear magnetoconductance stem from the same mechanism. \cred{In Ref.~\cite{supplement} we show that  the magnitude of $\gamma(B,\theta)/B$ is given by $n_{\mathrm{mg}}L_x\rho_{\mathrm{mg}} a g_{\mathrm{mg}}\mu_B hG(0)/e^2\times h G^2(0) U/eT^2$, where the temperature is measured in energy units, $n_{\mathrm{mg}}$ is the two-dimensional spatial density of midgap states, $\rho_{\mathrm{mg}}$ is the density of the midgap states in the energy space, $L_x$ is the length of the edge channel, and $a$ is the decay length of the midgap state wave function. The rest of quantities are defined in the main text. Then from the results of Ref.~\cite{chen2023linearG} on the linear magnetotransport due to midgap states we can conclude that $n_{\mathrm{mg}}L_x\rho_{\mathrm{mg}} a g_{\mathrm{mg}}\mu_B hG(0)/e^2\sim 10^{-1}\mathrm{T}^{-1}$. Here $G(0)$ is the zero-field linear edge conductance, $G(0)\approx 7.5\times 10^{-6}\mathrm{S}$, and the value for $L_x$ is roughly $L_x\sim 1\mu\mathrm{m}$.} From comparing the fits for linear and nonlinear magnetotransport, we obtain $h G^2(0) U/eT^2\sim 10^{-4}\mathrm{A\,V^{-2}}$ at temperature roughly equal to $10\,\mathrm{K}$. Then we obtain $U\sim 1\,K$, which is a very rough estimate.

The obtained value of $U$ is both encouraging and concerning. It is encouraging as it implies that the condition $U\lesssim \Gamma, T$ is indeed roughly satisfied for strongly scattering midgap states, and the presented analysis is self-consistent. Yet it is concerning because this is quite a small value for a Coulomb energy scale associated with a localized state. This value can be reconciled with physical reality if one assumes that the midgap states considered in Ref.~\cite{chen2023linearG} and in this work are in fact mesoscopic compressible regions, perhaps the `puddles' described in previous works on the lack of quantization of the zero-field conductance~\cite{vayrynen2013puddles,vayrynen2014prb}. The puddles must be large enough and polarizable enough to have a very small charging energy scale associated with them, yet are small enough that the mean level spacing for each such puddle is large compared to energy scales associated with the magnetic field, level width, or temperature. In other words, the bulk of the system should be a very poor conductor. These considerations require further explorations. 

\cred{Taken together, our previous works on magnetotransport in monolayer $\mathrm{WTe}_2$, Ref.~\cite{zhao2021determination,chen2023linearG}, and the present paper imply that small-$B$  and large-$B$  magnetotransport originate from distinct mechanisms. At higher fields, $B\gtrsim 2\,\mathrm T$, scattering occurs primarily on the edge. Accordingly, the nonlinear magnetoconductance stems from the current-induced spin polarization on the edge, and the associated effective Zeeman field due to electron-electron interaction~\cite{zhao2021determination,rudner2019currentgap}. At smaller fields, $B\lesssim 0.5\,\mathrm T$, scattering occurs away from the edge, on bulk midgap states. In this case the nonlinear magnetoconductance involves tunneling of the spin-polarized edge carriers onto a resonant impurity, and their precession in the combined Zeeman field due to external magnetic field, and the effective Zeeman field on the impurity site. The transition in the nature of the scattering mechanism is plausible, since at small fields the effect of potential scattering on the edge roughly scales as $B^2/E_F^2$, where $E_F$ is the edge Fermi energy counted from the position of the edge Dirac point, which determines the spin splitting at the Fermi level. At small $B$, the resistance due to scattering on the edge is weak as compared to the linear-in-$B$ scattering coming from the bulk states, but it starts to dominate at larger magnetic fields. }

The work of DAP was supported by the National Science Foundation Grant No. DMR-2138008. The experimental work was carried out as part of Programmable Quantum Materials, an Energy Frontier Research Center funded by the U.S. Department of Energy (DOE), Office of Science, Basic Energy Sciences (BES), under Award No. DESC0019443. Fabrication and measurement was performed using facilities and instrumentation supported by the U.S. National Science Foundation through the UW Molecular Engineering Materials Center (MEM-C), a Materials Research Science and Engineering Center (DMR-2308979).

\bibliography{Citation.bib}
\bibliographystyle{apsrev4-2}


\end{document}


\title{Supplement for Magnetochiral anisotropy on a quantum spin Hall edge}
\author{Youjian Chen}
\affiliation{Department of Physics, University of Virginia, Charlottesville, Virginia 22904, USA}
\author{Gary Quaresima}
\affiliation{Department of Physics, University of Virginia, Charlottesville, Virginia 22904, USA}
\author{Wenjin Zhao}
\affiliation{Kavli Institute at Cornell for Nanoscale Science, Ithaca, New York 14853, USA}
\author{Elliott Runburg}
\affiliation{Department of Physics, University of Washington, Seattle, Washington 98195 USA}
\author{David Cobden}
\affiliation{Department of Physics, University of Washington, Seattle, Washington 98195 USA}
\author{D. A. Pesin}
\affiliation{Department of Physics, University of Virginia, Charlottesville, VA 22904, USA}

\maketitle
\tableofcontents

\section{Hamiltonian of the problem}\label{suppsec:magnetizationofmifgapstate}

The Hamiltonian used in this work is represented by the interacting Anderson model~\cite{anderson1961model} containing four parts\cite{chen2023linearG}:
\begin{equation}\label{eq:totalH}
    H=H_{\mathrm{edge}}+H_{\mathrm{hyb}}+H_{\mathrm{mg}}+H_U.
\end{equation}
Here, $H_{\mathrm{edge}}$ is the Hamiltonian for the helical edge state electrons which can be expressed in the second-quantized form as
\begin{align}\label{eq:edgeH}
    H_{\mathrm{edge}}=\sum_{k,\sigma}\sigma v k a^\dagger_{k,\sigma}a_{k,\sigma}.
\end{align}
In the above equation $a_{k,\sigma}$ is the annihilation operator for the helical edge electrons with momentum $k$ along the edge direction, Dirac velocity $v$, and spin projections  $\sigma=\uparrow,\downarrow$ on the edge spin polarization axis. For small magnetic field, i.e, $B\lesssim 0.5T$, the Zeeman coupling for the helical edge electrons can be ignored, so we omit it in the Hamiltonian for the edge states. We assume that spin-up electrons propagate from left to right, and spin-down electrons propagate from right to left. 

Hamiltonian $H_{\mathrm{mg}}$ describes a midgap level Zeeman-coupled to magnetic field:
\begin{equation}\label{eq:ZeemanH}
    H_{\mathrm{mg}}=\begin{pmatrix}
     c^{\dagger}_{\uparrow} &
    c^{\dagger}_{\downarrow} \\
    \end{pmatrix} 
    \begin{pmatrix}
    \myeta+b_{z} & b_{x}-i b_{y} \\
    b_{x}+i b_{y} & \myeta-b_{z} \\
    \end{pmatrix}
    \begin{pmatrix}
    c_{\uparrow} \\
    c_{\downarrow}
    \end{pmatrix}.
\end{equation}
Here, $\epsilon_{\mathrm{mg}}$ is the energy of the midgap level, $\bm b \equiv(b_{x},b_{y},b_{z})$ is the Zeeman field. It is related to the external magnetic field $\bm B$ by $\bm{b}=\frac12 g_{\mathrm{mg}}\mu_{B}\bm{B}$, where $g_{\mathrm{mg}}$ is the $g$-factor for the midgap state and $\mu_{B}$ is the Bohr magneton. In this Supplementary Material, we will solve the problem in spherical coordinates, i.e, $\bm{b}\equiv(b \sin \theta \cos \phi, b \sin \theta \sin \phi, b \cos \theta)$, where $\theta$ is the azimuthal angle between the edge spin polarization and the Zeeman field, and $\phi$ is the polar angle with 'up' direction of the edge spin polarization as the z-axis.

In turn, $H_{\mathrm{hyb}}$ describes the hybridization between the helical edge electrons and a midgap state localized at the origin, $x_{\mathrm{mg}}=0$, where $x_{\mathrm{mg}}$ is the coordinate along the edge:
\begin{equation}\label{eq:hybH}
    H_{\mathrm{hyb}}= \sum_{k,\sigma}( t a^{\dagger}_{k,\sigma} c_{\sigma}+t^{*}c_{\sigma}^{\dagger} a_{k,\sigma}),
\end{equation}
where $c_{\sigma}$ is the annihilation operator for an electron with spin $\sigma$ in the midgap state, and $t$ is the spin-conserving and momentum-independent hybridization matrix element. Although the hybridization matrix element may not be spin-conserving and momentum-independent, we build our simplest model for magneto-conductance with minimum number of parameters. The spin-conserving hybridization matrix elements assumption is reasonable for describing magneto-conductance of monolayer WTe$_2$ whose band structure, both bulk and edge, is characterized by a definite spin projection onto a certain axis in the mirror plane of the material~\cite{thomale2019prb,zhao2021determination,nandy2022kp,chen2023linearG}. In what follows we use the following notation for normalization: the sum over states in momentum space is $\sum_{k}=\frac{L_{x}}{2 \pi} \int dk$. We also define the level width as $\Gamma=\frac{L_{x}|t|^{2}}{2 v}$ whose inverse describes the time that electrons stay in the midgap state during the scattering process.

Finally, to describe the mechanism of magnetochiral anisotropy considered in this work, one must introduce the Coulomb interaction between electrons on the midgap level. We use the standard Hubbard form of this interaction: 
\begin{equation}\label{eq:CoulombH}
H_{U}=U c^{\dagger}_{\uparrow}c^{\dagger}_{\downarrow}c_{\downarrow}c_{\uparrow}.
\end{equation}
Here $U$ describes the strength of the Coulomb interaction. As discussed in the main text, comparison of theoretical results to the experimental data confirms that the Anderson model we employ is in the nonmangetic limit, $U\lesssim \Gamma$, at least for midgap levels that scatter the edge electrons most efficiently.

\section{Hartree-Fock treatment of the Hubbard Hamiltonian}

The Hamiltonian of the problem~\eqref{eq:totalH} is a sum of noninteracting part describing the edge and migdap states and their hybridization, and an interacting part, which we chose in the standard Hubbard form: 
\begin{align}\label{eq:free+U}
    H=H_{\mathrm{free}}(\myeta,\bm{b})+H_U. 
\end{align}
Naturally, the free part of the Hamiltonian contains the applied magnetic field and the single-particle microscopic parameters of the problem. We will see that at the Hartree-Fock level the interaction renormalizes some of these parameters.

Following the original work of Anderson, we will employ the Hartree-Fock treatment of the Hubbard Hamiltonian. This reduces it to an effective single-particle one (`MF' stands for `mean-field'), 
\begin{equation}\label{eq:meanfieldHU}
    H^{\rm{MF}}_{U}= U \begin{pmatrix}
     c^{\dagger}_{\uparrow} &
    c^{\dagger}_{\downarrow} \\
    \end{pmatrix} 
    \begin{pmatrix}
    \braket{c^{\dagger}_{\downarrow}c_{\downarrow}} & -\braket{c^{\dagger}_{\downarrow}c_{\uparrow}} \\
     -\braket{c^{\dagger}_{\uparrow}c_{\downarrow}} & \braket{c^{\dagger}_{\uparrow}c_{\uparrow}} \\
    \end{pmatrix}
    \begin{pmatrix}
    c_{\uparrow} \\
    c_{\downarrow}
    \end{pmatrix}.
\end{equation}
Here, $\braket{...}$ is the average with the statistical operator of the system, $\rho$, normalized to $\mathrm{Tr}\rho=1$:
\begin{align}
    \braket{c^{\dagger}_{\sigma}c_{\sigma'}} =\mathrm{Tr}(c^{\dagger}_{\sigma}c_{\sigma'}\rho).
\end{align}
We emphasize that at this point $H^{\rm{MF}}_{U}$ depends on an arbitrary statistical operator $\rho$.
Eventually, we will use one appropriate for a Landauer transport problem.

The mean-field decomposition of the Hubbard Hamiltonian allows to write the following sequence of equations for the mean-field approximation for the total Hamiltonian:
\begin{align}\label{suppeq:HamiltonianMF}
        H^{\rm{MF}} = &\sum_{k \sigma} \sigma v k a^{\dagger}_{k,\sigma}a_{k,\sigma}+\sum_{k,\sigma}( t a^{\dagger}_{k,\sigma} c_{\sigma}+t^* c_{\sigma}^{\dagger} a_{k,\sigma})\nonumber \\
        &+\sum_{\sigma \sigma'}\left[\myeta+(\bm b \cdot \bm{\sigma})\right]_{\sigma\sigma'}c^{\dagger}_{\sigma}c_{\sigma'}+
    U \begin{pmatrix}
     c^{\dagger}_{\uparrow} &
    c^{\dagger}_{\downarrow} \\
    \end{pmatrix} 
    \begin{pmatrix}
    \braket{c^{\dagger}_{\downarrow}c_{\downarrow}} & -\braket{c^{\dagger}_{\downarrow}c_{\uparrow}} \\
     -\braket{c^{\dagger}_{\uparrow}c_{\downarrow}} & \braket{c^{\dagger}_{\uparrow}c_{\uparrow}} \\
    \end{pmatrix}
    \begin{pmatrix}
    c_{\uparrow} \\
    c_{\downarrow}
    \end{pmatrix}\nonumber \\
    =& H_{\mathrm{free}}(\myeta_{\mathrm{eff}}, \bm{b}_{\mathrm{eff}}),
\end{align}
with
\begin{align}\label{eq:effective-quantities}
\myeta_{\mathrm{eff}}:=& \myeta+\frac{U}{2}( \braket{c^{\dagger}_{\downarrow}c_{\downarrow}}
+\braket{c^{\dagger}_{\uparrow}c_{\uparrow}}
),\nonumber \\
\bm{b}_{\mathrm{eff}}:=& \bm{b} -\frac{U}{2}\sum_{\sigma\sigma'}\bm \sigma_{\sigma\sigma'}\braket{c^{\dagger}_{\sigma} c_{\sigma'}}.
\end{align}

It is evident from Eqs.~\eqref{suppeq:HamiltonianMF} that the mean-field Hamiltonian, $H_{\mathrm{MF}}$, looks like the original Hamiltonian, Eq.~\eqref{eq:totalH}, without the Coulomb interaction. That is, it coincides with $H_{\mathrm{free}}$ from Eq.~\eqref{eq:free+U}, but with bare $\myeta$ and $\bm b$ replaced with effective quantities $\myeta_{\mathrm{eff}}$ and
$\bm b_{\mathrm{eff}}$. 

The effective energy of the midgap state and Zeeman coupling in Eq.~\eqref{eq:effective-quantities} depend on yet unspecified statistical operator $\rho$ for the electrons, and the corresponding single-particle density matrix, besides their trivial dependence on $\myeta$ and $\bm b$. Below we will choose $\rho$ to correspond to the Landauer transport problem for the edge electrons with the mean-field Hamiltonian containing $\myeta_{\mathrm{eff}}$ and $\bm b_{\mathrm{eff}}$. This means that in general Eqs.~\eqref{eq:effective-quantities} must be solved self-consistently, as is normally done in the Hartree-Fock approach. In what follows, we will adopt the ``single-shot" Hartree-Fock approximation, in which Eqs.~\eqref{eq:effective-quantities} are solve to first order in $U$, such that the self-consistency aspect is not relevant. This should be a reasonable approximation for $U\lesssim \Gamma$~\cite{anderson1961model}. It should yield qualitatively reasonable results even for $U\sim \Gamma$, and according to estimates coming from the comparison to the experimental data $U$ is indeed rather small, justifying the procedure described below.

\section{Solution of the Lippmann-Schwinger equation}\label{suppsec:scattering}

We now proceed to solve the scattering problem for
Hamiltonian $H_{\rm{free}}(\myeta,\bm{b})$, treating $H_{\mathrm{hyb}}$ as the scattering part. This was done in our previous work~\cite{chen2023linearG}, where we used the Lippmann-Schwinger equation to solve the problem, so in this Section we largely repeat that derivation for completeness. 

It is convenient to describe the midgap state using the eigenstates of $H_{\mathrm{mg}}$, modifying the hybridization accordingly. The single-particle Hamiltonian of the midgap state can be diagonalized as 
\begin{equation}
    {\cal{U}} \begin{pmatrix}
    \myeta+b_{z} & b_{x}-i b_{y} \\
    b_{x}+i b_{y} & \myeta-b_{z} \\
    \end{pmatrix}
    {\cal{U}}^{\dagger}=
    \begin{pmatrix}
\epsilon_{+} & 0\\
0 &\epsilon_{-} \\
    \end{pmatrix},
\end{equation}
where the eigenvalues are $\epsilon_{\pm}=\myeta\pm b$. The unitary matrix ${\cal{U}}$ diagonalizing the midgap state Hamiltonian also relates the annihilation operators for the spin states of the midgap level, $c_{\uparrow,\downarrow}$, to those of the eigenstates of the midgap level Hamiltonian, $c_\pm$. Its matrix elements can be written using the spherical angles defining the direction of the Zeeman field, such that $\bm b=b(\sin\theta\cos\phi,\sin\theta\sin\phi,\cos\theta)$. We have then
\begin{equation}
    \begin{pmatrix}
    c_{+} \\
    c_{-}
    \end{pmatrix}
    =
    {\cal{U}}
    \begin{pmatrix}
    c_{\uparrow} \\
    c_{\downarrow}
    \end{pmatrix}
    =
     \begin{pmatrix}
    {\cal{U}}_{+\uparrow} & {\cal{U}}_{+\downarrow}\\
    {\cal{U}}_{-\uparrow} & {\cal{U}}_{-\downarrow}\\
    \end{pmatrix}
     \begin{pmatrix}
    c_{\uparrow} \\
    c_{\downarrow}
    \end{pmatrix}
\equiv
\begin{pmatrix}
\cos \frac{\theta}{2} & \sin \frac{\theta}{2} e^{-i \phi }\\
    -\sin \frac{\theta}{2} e^{+i \phi} &\cos\frac{\theta}{2} \\
\end{pmatrix}
\begin{pmatrix}
    c_{\uparrow} \\
    c_{\downarrow}
    \end{pmatrix},
\end{equation}
and
\begin{equation}
    \begin{pmatrix}
    c_{\uparrow} \\
    c_{\downarrow}
    \end{pmatrix}
    =
    {\cal{U}}^{\dagger}
    \begin{pmatrix}
    c_{+} \\
    c_{-}
    \end{pmatrix}
    =
     \begin{pmatrix}
    {\cal{U}}^{\dagger}_{\uparrow+} & {\cal{U}}^{\dagger}_{\uparrow -}\\
    {\cal{U}}^{\dagger}_{\downarrow+} & {\cal{U}}^{\dagger}_{\downarrow-}\\
    \end{pmatrix}
     \begin{pmatrix}
    c_{+} \\
    c_{-}
    \end{pmatrix}
\equiv
\begin{pmatrix}
\cos \frac{\theta}{2} & -\sin \frac{\theta}{2} e^{-i \phi }\\
    \sin \frac{\theta}{2} e^{+i \phi} &\cos\frac{\theta}{2} \\
\end{pmatrix}
\begin{pmatrix}
    c_{+} \\
    c_{-}
    \end{pmatrix}.
\end{equation}
Here ${\cal{U}}_{s\sigma}$ and ${\cal{U}}^{\dagger}_{\sigma s}$ are the matrix elements of ${\cal{U}}$ and of its Hermitian conjugate, and $s$ is the index labeling mid-gap eigenstates. 

In the new basis, the midgap level Hamiltonian is written as 
\begin{equation}
    H_{\mathrm{mg}}=
    \begin{pmatrix}
    c^{\dagger}_{+} &
    c^{\dagger}_{-}
    \end{pmatrix}
    \begin{pmatrix}
\epsilon_{+} & 0\\
0 &\epsilon_{-} \\
    \end{pmatrix}
    \begin{pmatrix}
    c_{+} \\
    c_{-}
    \end{pmatrix},
\end{equation}
while the hybridization between the edge electrons with spins $\uparrow,\downarrow$ and the midgap level eigenstates labeled with $\pm$ becomes
\begin{equation}
    V_{\sigma s}\equiv  t\, {\cal{U}}^{\dagger}_{\sigma s}=
    \begin{pmatrix}
        V_{\uparrow+} & V_{\uparrow-}\\
         V_{\downarrow+} & V_{\downarrow-} \\
    \end{pmatrix}
    =
   t \begin{pmatrix}
        \cos\frac{\theta}{2} & -\sin\frac{\theta}{2}e^{-i\phi} \\
        \sin\frac{\theta}{2}e^{+i\phi} & \cos\frac{\theta}{2}\\
    \end{pmatrix},
\end{equation}
and
\begin{equation}
    V_{s \sigma}\equiv t^{*}\, {\cal{U}}_{s \sigma}=
    \begin{pmatrix}
        V_{+\uparrow} & V_{+\downarrow}\\
         V_{-\uparrow} & V_{-\downarrow} \\
    \end{pmatrix}
    =
   t^{*} \begin{pmatrix}
        \cos\frac{\theta}{2} & \sin\frac{\theta}{2}e^{-i\phi} \\
        -\sin\frac{\theta}{2}e^{+i\phi} & \cos\frac{\theta}{2}\\
    \end{pmatrix}.
\end{equation}
The hybridization Hamiltonian is now
    \begin{equation}
    H_{\mathrm{hyb}}=\sum_{k,\sigma=\uparrow,\downarrow}\sum_{s=+,-}V_{\sigma s} a_{k, \sigma}^{\dagger}c_{s}+V_{s \sigma} c_{s}^{\dagger}a_{k,\sigma}.
\end{equation}

In what follows we consider the single-particle retarded Green's function and the T-matrix for the scattering problem at hand. We first define a few quantities for the case of uncoupled edge and midgap level. The single-particle retarded Green's function can be written as the sum of single-particle operators acting in the the edge and midgap level subspaces of the full electronic Hilbert space that contains the edge and midgap states: 
\begin{equation}
    \hat{G}_{0}=\hat{G}_{\mathrm{edge}}+\hat{G}_{\mathrm{mg}}.
\end{equation}
The edge Green's function is written as
\begin{equation}\label{eq:edgeGF}
    \hat{G}_{\mathrm{edge}}=\sum_{k\sigma=\uparrow,\downarrow}\frac{\ket{k ,\sigma} \bra{k ,\sigma}}{E+i \eta - \sigma v k},
\end{equation}
where $\eta$ is a positive infinitesimal, and $\ket{k\sigma}$ are the single-particle eigenstates of $H_{\mathrm{edge}}$ with momentum $k$ and spin $\sigma$. The midgap level Green's function is
\begin{equation}
    \hat{G}_{\mathrm{mg}}=\sum_{s=\pm}\frac{\ket{s}\bra{s}}{E+ i\eta-\epsilon_{s}},
\end{equation}
where $\ket{s}$  with $s=\pm$ are the midgap states with energies $\epsilon_\pm$.

The hybridization between the edge and midgap states can be decomposed into a sum of two parts, $\hat{V}=\hat{\mathcal{V}}+\hat{\mathcal{V}}^{\dagger}$. The operator $\mathcal{V}$ only has matrix elements for transitions from the midgap state into edge states:
\begin{equation}
\hat{\mathcal{V}}=\sum_{k}\sum_{s,\sigma}V_{\sigma s}\ket{k,\sigma}\bra{s},
\end{equation}
while $\hat{\mathcal{V}}^{\dagger}$ only has matrix elements for transitions from the edge states to the midgap level:
\begin{equation}
\hat{\mathcal{V}}^{\dagger}=\sum_{k} \sum_{s,\sigma}V_{s\sigma } \ket{s}\bra{k,\sigma}.
\end{equation}
The equation for the T-matrix in the full electronic Hilbert space, $T_{\mathrm{full}}$, has the usual form of   $T_{\mathrm{full}}=\sum_{n=0}^{\infty}\hat{V}(\hat{G}_{0}\hat{V})^{n}$. We divide the full T-matrix into four parts:
\begin{equation}
T_{\mathrm{full}}=T_{\mathrm{mg}}+T_{\mathrm{edge}}+T_{\mathrm{mix}}+T_{\mathrm{mix}}^{\dagger}.
\end{equation}
Here $T_{\mathrm{mg}}$ describes the scattering process between mid-gap states:
\begin{equation}
T_{\mathrm{mg}}=\sum_{s s'}T_{\mathrm{mg},ss'}\ket{s}\bra{s'}.
\end{equation}
Where $T_{\mathrm{mg},ss'}=\bra{s} T_{\mathrm{full}} \ket{s'}$  describes the scattering matrix elements between mid-gap states $\ket{s}$ and $\ket{s'}$.

$T_{\mathrm{edge}}$ describes the scattering process between edge states:
\begin{equation}
T_{\mathrm{edge}}=\sum_{k k'} \sum_{\sigma \sigma'} T_{\mathrm{edge},\sigma \sigma'} \ket{k,\sigma}\bra{k',\sigma'}.
\end{equation}
Here $T_{\mathrm{edge},\sigma\sigma'}=\bra{k\sigma} T_{\mathrm{full}} \ket{k'\sigma'}$ is independent of $k$ and $k'$ since the hybridization is independent of momentum.

$T_{\mathrm{mix}}$ describes the scattering process from edge states to mid-gap states:
\begin{equation}
T_{\mathrm{mix}}=\sum_{k} \sum_{s \sigma} T_{\mathrm{mix},s\sigma} \ket{s}\bra{k,\sigma}.
\end{equation}
Here $T_{\mathrm{mix},s\sigma}=\bra{s} T_{\mathrm{full}} \ket{k,\sigma}$
is likewise independent of $k$.
$T_{\mathrm{mix}}$  describes the scattering matrix elements from edge states $\ket{k, \sigma}$ to mid-gap state $\ket{s}$ with  momentum $k$.

$T_{\mathrm{mix}}^{\dagger}$ describes the scattering process from mid-gap states to edge states, and is initially defined separately. However it will turn out in the Lippmann-Schwinger equation solution, that is it simply the hermitian conjugation of $T_{\mathrm{mix}}$.

We can get the perturbation series for $T_{\mathrm{edge}}$ and $T_{\mathrm{mix}}$
\begin{equation}
T_{\mathrm{edge}}=\sum_{n=0}^{+\infty}\hat{\mathcal{V}}\hat{G}_{\mathrm{mg}}\hat{\mathcal{V}}^{\dagger}(\hat{G}_{\mathrm{edge}}\hat{\mathcal{V}}\hat{G}_{\mathrm{mg}}\hat{\mathcal{V}}^{\dagger})^{n} ,
\end{equation}
and
\begin{equation}
    T_{\mathrm{mix}}=\sum_{n=0}^{+\infty}\hat{\mathcal{V}}^{\dagger} (\hat{G}_{\mathrm{edge}}\hat{\mathcal{V}}\hat{G}_{\mathrm{mg}}\hat{\mathcal{V}}^{\dagger})^{n}.
\end{equation}
The perturbation series for $T_{\mathrm{edge}}$ and $T_{\mathrm{mix}}$ can be simplified to
\begin{equation}\label{Lippmannschwingerforedge}
T_{\mathrm{edge}}-T_{\mathrm{edge}}\hat{G}_{\mathrm{edge}}\hat{\mathcal{V}}\hat{G}_{\mathrm{mg}}\hat{\mathcal{V}}^{\dagger}=\hat{\mathcal{V}}\hat{G}_{\mathrm{mg}}\hat{\mathcal{V}}^{\dagger}.
\end{equation}

The result of our previous work~\cite{chen2023linearG} for the T-matrix for the edge reads
\begin{equation}\label{eq:edgeTmatrix}
    T_{\mathrm{edge},\sigma\sigma'}\equiv\begin{pmatrix}
    T_{\mathrm{edge},\uparrow\uparrow} & T_{\mathrm{edge},\uparrow\downarrow} \\
    T_{\mathrm{edge},\downarrow\uparrow} & T_{\mathrm{edge},\downarrow\downarrow}
    \end{pmatrix}
    =
    \frac{|t|^{2}}{(E-\myeta+i \Gamma)^{2}-b^{2}}
    \begin{pmatrix}
    E-\myeta+i\Gamma+b \cos \theta & b \sin \theta e^{-i\phi} \\
    b \sin \theta e^{+i\phi} & E-\myeta+i\Gamma-b \cos \theta
    \end{pmatrix}.
\end{equation}
The T-matrix for edge-mid gap scattering $T_{\mathrm{mix}}$ is obtained from
\begin{equation}   T_{\mathrm{mix}}=\hat{\mathcal{V}}^{\dagger}+\hat{\mathcal{V}}^{\dagger}\hat{G}_{\mathrm{edge}}T_{\mathrm{edge}}.
\end{equation}
The algebraic equation for $T_{\mathrm{mix}}$ matrix elements is
\begin{equation}
    T_{\mathrm{mix},s\sigma}=V_{s\sigma}-\frac{iL_{x}}{2 v}\sum_{\sigma'}V_{s\sigma'}T_{\mathrm{edge},\sigma' \sigma}=\sum_{\sigma'}V_{s\sigma'}(\delta_{\sigma'\sigma}-\frac{i L_{x}}{2v}T_{\mathrm{edge},\sigma' \sigma}).
\end{equation}

For a given energy $E$, the system has two eigenstates labeled by quantum number $\sigma=\uparrow$ or $\downarrow$ indicating that each eigenstate contains an incoming wave with spin $\sigma$. The eigenstate with quantum number $\sigma$ and energy $E$ is related to the edge state with spin $\sigma$ momentum $k$ by:
\begin{equation}  \ket{\sigma,E}=\ket{k,\sigma}+G_{\mathrm{edge}}T_{\mathrm{edge}}\ket{k,\sigma}+G_{\mathrm{mg}}T_{\mathrm{mix}}\ket{k,\sigma}.
\end{equation}
For electrons coming from the left with energy $E$, the incoming state is $\ket{k,\uparrow}$, so the scattering state is
\begin{equation}
\ket{\uparrow,E}=\ket{k,\uparrow}+G_{\mathrm{edge}}T_{\mathrm{edge}}\ket{k,\uparrow}+G_{\mathrm{mg}}T_{\mathrm{mix}}\ket{k,\uparrow}.
\end{equation}
For electrons coming from the right with energy $E$, the incoming state is $\ket{k,\downarrow}$, and the scattering state is
\begin{equation}  
\ket{\downarrow,E}=\ket{k,\downarrow}+G_{\mathrm{edge}}T_{\mathrm{edge}}\ket{k,\downarrow}+G_{\mathrm{mg}}T_{\mathrm{mix}}\ket{k,\downarrow}.
\end{equation}
Note that the solution to the Lippmann-Schwinger equation maintains its normalization, so it follows $\braket{\sigma, E|\sigma',E'}=\delta_{\sigma \sigma'}\delta(E-E')$.

Using the relation between mid-gap state spin bra  $\bra{\sigma}$ and diagonal basis bra $\bra{s}$, 
\begin{equation}
    \bra{\sigma}= \sum_{s}{\cal{U}}_{\sigma s}^{\dagger} \bra{s},
\end{equation}
we can obtain
\begin{align}
    \braket{\sigma_{1}|\sigma_{2},E}=\sum_{s}{\cal{U}}^{\dagger}_{\sigma_{1} s}  \bra{s}G_{\mathrm{mg}} T_{\mathrm{mix}} \ket{k,\sigma_{2}}.
\end{align}
The algebraic equation for the overlap between mid-gap states in spin basis $\bra{\sigma_{1}}$ and the energy eigenstate $\ket{\sigma_{2},E}$
\begin{equation}
  \braket{\sigma_{1}|\sigma_{2},E}= \sum_{s}\frac{{\cal{U}}^{\dagger}_{\sigma_{1} s}T_{s \sigma_{2}}}{E+i\eta-\epsilon_{s}}=\sum_{s} \sum_{\sigma'}\frac{t^{*}{\cal{U}}^{\dagger}_{\sigma_{1} s}{\cal{U}}_{s\sigma'}}{E+i\eta-\epsilon_{s}}(\delta_{\sigma'\sigma_{2}}-\frac{i L_{x}}{2v}T_{\mathrm{edge},\sigma' \sigma_{2}}).   
\end{equation}
We can define matrix $M(E)$ proportional to the overlap of mid-gap state  and energy eigenstate as $\braket{\sigma_{1}|\sigma_{2},E}=t^{*} M_{\sigma_{1}\sigma_{2}}$
\begin{equation}
    M(E)=
    \begin{pmatrix}
      M_{\uparrow \uparrow}(E)  & M_{\uparrow \downarrow}(E) \\
      M_{\downarrow \uparrow}(E) & M_{\downarrow \downarrow}(E) \\
    \end{pmatrix}
    =\frac{1}{(E-\myeta + i \Gamma)^2-b^2}
    \begin{pmatrix}
        E-\myeta + i \Gamma +b \cos{\theta} & b \sin{\theta}e^{-i \phi} \\
        b \sin{\theta}e^{+i \phi} & E -\myeta + i \Gamma -b \cos{\theta}\\
    \end{pmatrix}.
\end{equation}
and its Hermitian conjugate $M^{\dagger}(E)$, related to the overlap of mid-gap state and energy eigenstate by $\braket{\sigma_{2},E|\sigma_{1}}=t M^{\dagger}_{\sigma_{1}\sigma_{2}}$
\begin{equation}
    M^{\dagger}(E)=
     \begin{pmatrix}
      M^{\dagger}_{\uparrow \uparrow}(E)  & M^{\dagger}_{\uparrow \downarrow}(E) \\
      M^{\dagger}_{\downarrow \uparrow}(E) & M^{\dagger}_{\downarrow \downarrow}(E) \\
    \end{pmatrix}
    =\frac{1}{(E-\myeta - i \Gamma)^2-b^2}
    \begin{pmatrix}
        E-\myeta - i \Gamma +b \cos{\theta} & b \sin{\theta}e^{-i \phi} \\
        b \sin{\theta}e^{+i \phi} & E -\myeta - i \Gamma -b \cos{\theta}\\
    \end{pmatrix}.
\end{equation}

\section{Nonequilibrium corrections to the Hartree-Fock Hamiltonian}\label{suppsec:correlators}

In this Section we use the results of the preceding one to evaluate $\braket{c^{\dagger}_{\sigma_{1}}c_{\sigma_{2}}}$ - the equal time correlators for spin $\sigma_{1}$ and $\sigma_{2}$ electrons on the impurity site - and then use it to evaluate the interaction corrections to the on-site energy of the midgap level and the exchange field. 

\subsection{Statistical operator for the electrons}

In Section~\ref{suppsec:scattering} we solved the single-particle part of the problem, determining the scattering states that diagonalize the mean-field Hamiltonian. We recall that these states belong to the continuous spectrum, and can be labeled with the motion direction of the incident wave. On a helical edge, this direction is uniquely determined by the spin of the incident electron, such that the annihilation operators for the scattering states can be labeled as $d_{\sigma,E}$. In terms of these operators, the mean field Hamiltonian is written as $H_{\mathrm{MF}}=H_\uparrow+H_\downarrow$, where
\begin{align}
    H_\sigma=\int dE \, E d^\dagger_{\sigma,E}d_{\sigma,E}.
\end{align}
The entire information about the scattering states is contained in the single-particle states to which $d_{\sigma,E}$ pertain. 

The central assumption of the Landauer transport theory is that the occupation probabilities for the scattering states are enforced by the leads from which the incident waves are emanating. Here we make an assumption that the midgap states relevant for scattering are sparse enough for the incident electrons to have equilibrium distribution for each helical branch, with the same temperatures, but unequal chemical potentials. That is, the statistical operator for the electrons is 
\begin{align}\label{eq:statoperator}
    \rho=e^{\frac{-1}{T}\left(H_\uparrow-\mu_L N_\uparrow\right)}\;
    e^{\frac{-1}{T}\left(H_\downarrow-\mu_R N_\downarrow\right)}. 
\end{align}
Here $N_{\uparrow,\downarrow}$ are operators of numbers of electrons incident from the left and right, labeled with their spin. 

The crucial simplification of using the scattering states is that the statistical averages for the corresponding operators with statistical operator~\eqref{eq:statoperator} read 
\begin{equation}\label{eq:exactcorrelators}
    \begin{pmatrix}
        \braket{d^{\dagger}_{\uparrow,E}d_{\uparrow,E'}} &  \braket{d^{\dagger}_{\uparrow,E}d_{\downarrow,E'}} \\
         \braket{d^{\dagger}_{\downarrow,E}d_{\uparrow,E'}} & \braket{d^{\dagger}_{\downarrow,E}d_{\downarrow,E'}}\\
    \end{pmatrix}
    =
    \begin{pmatrix}
        n_{L}(E) & 0 \\
        0 & n_{R}(E)  \\
    \end{pmatrix}\delta(E-E').
\end{equation}
In the above equation the occupations for the scattering states, $n_{L,R}(E)$, are given by the Fermi-Dirac distribution functions with the same temperature $T$, but different chemical potentials $\mu_{L,R}$, respectively:
\begin{align}
    n_{L,R}(E)=n_F\left(E-\mu_{L,R}\right) = \left[\exp\left(\frac{E-\mu_{L,R}}{T}\right)+1\right]^{-1} .
\end{align}

\subsection{Correlators for the midgap level}

To evaluate $\braket{c^{\dagger}_{\sigma_{1}}c_{\sigma_{2}}}$ using Eq.~\eqref{eq:exactcorrelators} we wish to express $c_{\sigma}$ via the annihilation operators for the exact scattering states, $d_{\sigma,E}$. We can accomplish this using the invariance of the field operators with respect to a choice of the single particle basis. Assuming that the edge states and the midgap level states form a complete orthonormal basis, \textit{i.e.} discarding the bulk band states as being too far in energy to affect dynamics, we can write for the field operator:
\begin{equation}\label{eq:FieldOp}
    \psi_{\sigma}(x)=\sum_{s=\pm} \braket{x,\sigma|s} c_{s}+\sum_{k,\sigma'}\braket{x,\sigma|k,\sigma'}a_{k,\sigma'}=\sum_{\sigma',E}\braket{x,\sigma|\sigma',E} d_{\sigma',E},
\end{equation}
\begin{equation}\label{eq:FieldOpdagger}
    \psi_{\sigma}^{\dagger}(x)=\sum_{s=\pm} \braket{s|x,\sigma}c^{\dagger}_{s}+\sum_{k,\sigma'}\braket{k,\sigma'|x,\sigma}a^{\dagger}_{k,\sigma'}=\sum_{\sigma',E}\braket{\sigma',E|x,\sigma} d^{\dagger}_{\sigma',E}.
\end{equation}
Here $\braket{x,\sigma|s}$ and $\braket{s|x,\sigma}$ are spin $\sigma$ components of the real-space wavefunctions of the mid-gap states with quantum number $s$, and its conjugate; $\braket{x ,\sigma|k,\sigma'}$ and $\braket{k,\sigma'|x,\sigma}$ are the wavefunction of helical edge with spin $\sigma'$ and momentum $k$ in real space, and its conjugate; $\braket{x,\sigma|\sigma',E}$ and $\braket{\sigma',E|x,\sigma}$ are spin $\sigma$ components of wavefunction of the system with quantum number $\sigma'$ and energy $E$ in real space and its conjugate. We used the following implied correspondence between sums and and integrals: $\sum_{k}=\frac{L_{x}}{2 \pi} \int dk= \frac{L_{x}}{2 \pi v} \int dE = \sum_{E}$

We set the commutation relations for the creation and annihilation operator of midgap states and helical edge states to be the conventional ones:
\begin{equation}
    \{c_{s},c^{\dagger}_{s'} \}=\delta_{s s'},
\end{equation}
and
\begin{equation}
    \{a_{k,\sigma},a^{\dagger}_{k',\sigma'} \}= \delta_{k k'}\delta_{\sigma \sigma'},
\end{equation}
which corresponds to the normalized wave function for helical edge states defined as 
\begin{equation}
  \braket{x ,\sigma|k,\sigma'}=\frac{\delta_{\sigma \sigma'}}{\sqrt{L_{x}}}e^{ i k x}.
\end{equation}
The commutation relations for field operators are
\begin{equation}
    \{\psi_{\sigma}(x),\psi^{\dagger}_{\sigma'}(x') \}=\delta_{\sigma \sigma'}\delta(x-x'), 
\end{equation}
while the commutation relations for the creation and anihilation operators for the scattering states are 
\begin{equation}
     \{d_{\sigma,E},d^{\dagger}_{\sigma',E'} \}=  \delta_{\sigma \sigma'}\delta(E-E').
\end{equation}

The occupation of energy eigenstates are
\begin{equation}
    \begin{pmatrix}
        \braket{d^{\dagger}_{\uparrow,E}d_{\uparrow,E'}} &  \braket{d^{\dagger}_{\uparrow,E}d_{\downarrow,E'}} \\
         \braket{d^{\dagger}_{\downarrow,E}d_{\uparrow,E'}} & \braket{d^{\dagger}_{\downarrow,E}d_{\downarrow,E'}}\\
    \end{pmatrix}
    =
    \begin{pmatrix}
        n_{L}(E) & 0 \\
        0 & n_{R}(E)  \\
    \end{pmatrix}\delta(E-E').
\end{equation}

The annihilation and creation operator for energy mid-gap state $c_{\sigma}$ and $c_{\sigma}^{\dagger}$ can be expressed 
\begin{equation}
c_{\sigma}=\sum_{\sigma'}\sum_{E} \braket{\sigma|\sigma',E}d_{\sigma',E}
\end{equation}
by integrating equation \eqref{eq:FieldOp}. Similarly,
\begin{equation}
c_{\sigma}^{\dagger}=\sum_{\sigma'}\sum_{E} \braket{\sigma',E|\sigma}d_{\sigma',E}^{\dagger}.
\end{equation}
So the correlator $\braket{c^{\dagger}_{\sigma_{1}}c_{\sigma_{2}}}$ is
\begin{equation}    \braket{c^{\dagger}_{\sigma_{1}}c_{\sigma_{2}}}=\sum_{\sigma}\sum_{E}\sum_{\sigma'}\sum_{E'} \braket{\sigma,E|\sigma_{1}} \braket{d^{\dagger}_{\sigma,E}d_{\sigma',E'}} \braket{\sigma_{2}|\sigma',E'}.
\end{equation}
This equation can be simplified to
\begin{align}\label{correlator}
\braket{c^{\dagger}_{\sigma_{1}}c_{\sigma_{2}}}=&\frac{\Gamma}{2 \pi}  \int d E (M(E)M^{\dagger}(E))_{\sigma_{2}\sigma_{1}}(E)(\nL{E}+\nR{E}) +\frac{\Gamma}{2 \pi}  \int d E (M(E)\sigma_{z}M^{\dagger}(E))_{\sigma_{2}\sigma_{1}}(E)(\nL{E} -\nR{E})
\\
=&:\braket{c^{\dagger}_{\sigma_{1}}c_{\sigma_{2}}}_{{\mathrm{eq}}} +
\braket{c^{\dagger}_{\sigma_{1}}c_{\sigma_{2}}}_{{\mathrm{neq}}},
\end{align}
where as a shorthand we define the first term as the ``equilibrium part" and the second term as the ``nonequilibrium part". The latter exists onlly in the presence of a transport current. The former is finite even without a transport current, but can still receive nonequilibrium corrections due to bias-induced changes to the density of electrons (admittedly, this makes ``equilibrium"  a bit of a misnomer for this quantity). 

Finally, the explicit form of the matrices in the integrands of Eq.~\eqref{correlator} is as follows:
\begin{align}
   M(E)M^{\dagger}(E) &= \frac{1}{(E-\myeta-b+i \Gamma)(E-\myeta+b+i \Gamma)(E-\myeta-b-i \Gamma)(E-\myeta+b-i \Gamma)} \nonumber\\
   & \begin{pmatrix}
        (E-\myeta+ b \cos{\theta})^2+\Gamma^2+ b^2 \sin^2{\theta} & 2 (E-\myeta) b \sin{\theta}e^{-i \phi} \\
        2 (E-\myeta) b \sin{\theta} e^{+i \phi} & (E-\myeta- b \cos{\theta})^2+\Gamma^2+ b^2 \sin^2{\theta} \\
    \end{pmatrix},
\end{align}
\begin{align}
    M(E)\sigma_{z}M^{\dagger}(E) &= \frac{1}{(E-\myeta-b+i \Gamma)(E-\myeta+b+i \Gamma)(E-\myeta-b-i \Gamma)(E-\myeta+b-i \Gamma)} \nonumber\\
    & \begin{pmatrix}
        (E-\myeta+ b \cos{\theta})^2+\Gamma^2- b^2 \sin^2{\theta} & 2(i \Gamma+b \cos{\theta})b \sin{\theta}e^{-i \phi} \\
        2(-i \Gamma+b \cos{\theta})b \sin{\theta}e^{+i \phi} & -(E-\myeta- b \cos{\theta})^2-\Gamma^2+ b^2 \sin^2{\theta}\\
    \end{pmatrix}.
\end{align}

\subsection{Corrections to the midgap state energy and the Zeeman field}\label{sec:Hfcorrections}



The decomposition of the correlators into equilibrium and nonequilibrium parts carries over to the corrections of to the 
Zeeman field and the energy of the midgap state.
Note that the equilibrium part is related to the sum of distribution functions on the edges attached to left and right sources with chemical potential $\mu_{L}$ and $\mu_{R}$, while the nonequilibrium part that is related to the difference of distribution functions on the edges attached to left and right sources with chemical potential $\mu_{L}$ and $\mu_{R}$.

The correction to the Zeeman field and the midgap energy are related to the correlators by: 
\begin{align}\label{seq:effective-parameters}
         &\delta b_{x}=-U \mathrm{Re} \braket{c^{\dagger}_{\uparrow} c_\downarrow}\nonumber\\
         &\delta b_{y}=-U \mathrm{Im} \braket{c^{\dagger}_{\uparrow} c_\downarrow}\nonumber \\
         &\delta b_{z}=\frac{U}{2}(\braket{c^{\dagger}_{\downarrow} c_\downarrow}-\braket{c^{\dagger}_{\uparrow} c_\uparrow})\nonumber \\
        &\delta \myeta=\frac{U}{2}(\braket{c^{\dagger}_{\downarrow} c_\downarrow}+\braket{c^{\dagger}_{\uparrow} c_\uparrow}).
\end{align}
Here $\delta b_{i},i=x,y,z$ are the components of correction to the Zeeman field $\delta \bm b$. We simply pattern match
\begin{align}\label{seq:effective-parameters}
         &\delta b_{x,\rm{(n)eq}}=-U \mathrm{Re} \braket{c^{\dagger}_{\uparrow} c_\downarrow}_{\rm{(n)eq}}\nonumber\\
         &\delta b_{y,\rm{(n)eq}}=-U \mathrm{Im} \braket{c^{\dagger}_{\uparrow} c_\downarrow}_{\rm{(n)eq}} \nonumber\\
         &\delta b_{z,\rm{(n)eq}}=\frac{U}{2}\left(\braket{c^{\dagger}_{\downarrow} c_\downarrow}_{\rm{(n)eq}}-\braket{c^{\dagger}_{\uparrow} c_\uparrow}_{\rm{(n)eq}}\right) \nonumber\\
        &\delta \myeta_{\rm{(n)eq}}=\frac{U}{2}\left(\braket{c^{\dagger}_{\downarrow} c_\downarrow}_{\rm{(n)eq}}+\braket{c^{\dagger}_{\uparrow} c_\uparrow}_{\rm{(n)eq}}\right)
\end{align}
to make the decomposition
\begin{align}
        &\delta \bm{b} = \delta \bm{b}_{\rm{\mathrm{eq}}} + \delta \bm{b}_{\rm{\mathrm{neq}}},\nonumber\\
        &\delta \myeta = \delta \myeta_{\rm{\mathrm{eq}}} + \delta \myeta_{\rm{\mathrm{neq}}}.\nonumber
\end{align}

After some trivial transformations, we obtain the following integrals for the equilibrium parts of the effective quantities: 
\begin{equation}\label{eq:equilibriumepsilon}
    \delta \myeta_{\mathrm{eq}}=-\frac{U \Gamma}{2\pi}\int_{-\infty}^{+\infty}dE\,\frac{\left((E-\myeta)^{2}+\Gamma^2+ b^{2}\right)\left(\nL{E}+\nR{E}\right)}{\left((E-\myeta-b)^2+\Gamma^2\right)\left((E-\myeta+b)^2+\Gamma^2\right)},
\end{equation}
and 
\begin{equation}\label{eq:equilibriumb}
    \delta \bm{b}_{\rm{\mathrm{eq}}} = -\bm{b} \cdot
    \frac{U \Gamma}{\pi}\int_{-\infty}^{+\infty}dE\,\frac{\left( E-\myeta \right)\left(\nL{E}+\nR{E}\right)}{\left((E-\myeta-b)^2+\Gamma^2\right)\left((E-\myeta+b)^2+\Gamma^2\right)}.
\end{equation}

For the nonequilibrium counterparts of these quantities we obtain 

\begin{equation}\label{eq:nonequilibriumepsilon}
    \delta \myeta_{\mathrm{neq}}=\frac{U \Gamma}{\pi}\int_{-\infty}^{+\infty}dE\,\frac{\left((E-\myeta) b \cos \theta\right)\left(\nL{E}-\nR{E}\right)}{\left((E-\myeta-b)^2+\Gamma^2\right)\left((E-\myeta+b)^2+\Gamma^2\right)},
\end{equation}
and 

\begin{align}\label{eq:nonequilibriumb}
\begin{pmatrix}
    \delta b_{x,\mathrm{neq}}\\
    \delta b_{y,\mathrm{neq}}\\
    \delta b_{z,\mathrm{neq}}
\end{pmatrix}
=-\frac{U \Gamma}{\pi}\int_{-\infty}^{+\infty}dE\,\frac{\left(\nL{E}-\nR{E}\right)}{\left((E-\myeta-b)^2+\Gamma^2\right)\left((E-\myeta+b)^2+\Gamma^2\right)}
\begin{pmatrix}
b^2 \cos \theta \sin \theta \cos \phi+\Gamma b \sin \theta \sin \phi\\
    b^2 \cos \theta \sin \theta \sin \phi-\Gamma b \sin \theta \cos \phi\\
    \frac12\left[(E-\myeta)^2+\Gamma^2+b^2 \cos 2 \theta\right]
\end{pmatrix}.
\end{align}

From Eqs.~\eqref{eq:equilibriumepsilon} and~\eqref{eq:equilibriumb} we observe that the ``equilibrium" quantities cannot lead to any odd-in-B dependence of the nonlinear edge conductance, despite their dependence on the applied bias across an impurity. Specifically, $\delta \myeta_{\mathrm{eq}}$ represents a shift the position of the midgap level, which is even in the magnetic field, and is clearly irrelevant for the antisymmetric conductance. In turn, $\delta \bm{b}_{\mathrm{eq}}$ is a shift in the magnetic field that is strictly odd in the applied one, hence the total mean-field B-field would also be strictly odd in the applied one. This implies that the Landauer conductance, calculated with this total field, is also even in the external field, despite being non-linear in the sense that it depends on the applied voltage. The equilibrium corrections will be discarded from now on.

Instead, the nonequilibrium quantities in Eqs.~\eqref{eq:nonequilibriumepsilon} and~\eqref{eq:nonequilibriumb} do lead to an antisymmetric part of the conductance. Under B-field reversal operation, effected by $\theta\to \pi-\theta,\,\phi\to\phi+\pi$, $\delta \myeta_{\mathrm{neq}}$ is odd, while $\delta \bm{b}_{\mathrm{neq}}$ contains an even part. It is then obvious that magnetic field reversal changes transmission probability through a midgap level, and this change depends on the applied current. Hence $\delta \myeta_{\mathrm{neq}}$ and the even-in-B part of $\delta \bm{b}_{\mathrm{neq}}$  describe magnetochiral anisotropy. 

\section{Observables and averaging over disorder}\label{suppsec:disorderaverage}

In the preceding Section we have solved the scattering problem for a single midgap state with on-site interaction in the Hartree-Fock approximation. Below we will use these results to first discuss the linear and nonlinear resistances of a single impurity, and later perform disorder average for the realistic situation with many midgap levels in the sample.

\subsection{Linear and nonlinear resistance for a collection of midgap levels}

We start out discussion of physical observable associated with the scattering problem considered above with the case of a single midgap level. 

For a single midgap level characterized with on-site energy $\myeta$ and width $\Gamma$ the transmission, $\cal T$, and reflection, $\cal R$, probabilities at electronic energy $E$ are given by~\cite{chen2023linearG}
 \begin{align}
&{\cal T}(E,\myeta,\Gamma,\bm{b})=1-{\cal R}(E,\myeta,\Gamma,\bm b)=1-\frac{4\Gamma^2(\bm b\times \bm z)^2}
    {((E-\myeta)^2+\Gamma^2-\bm b^2)^2+4\Gamma^2\bm b^2}.
\end{align}

At a finite temperature $T$, it is convenient to introduce the energy-averaged transmission coefficient, $\bar{\cal T}$: 
\begin{equation}
    \bar{\cal T}(\myeta,\Gamma,\bm b)=\int dE\;{ \cal T}(E,\myeta,\bm b)\;\left[-\frac{\partial n_F\left(E-\mu\right)}{\partial E}\right],
\end{equation}
and an analogous equation for the average reflection coefficient, $\bar{\cal R}$. For $T\to 0$, these quantities reduce to the transmission and reflection probabilities at the Fermi level, $E=\mu$. 

One can define many conductances or resistances associated with the single-impurity problem. For instance, the standard linear two-point Landauer conductance is given by 
\begin{align}
G_{\rm{2pt}}(\myeta,\Gamma,\bm b)=\frac{e^2}{h}\bar{\cal T}(\myeta,\Gamma,\bm b). 
\end{align}

It is well-known~\cite{datta2005quantum} that for a collection of scatterers in the incoherent regime, the four-point linear ``intrinsic" resistance of each of them,
\begin{align}
    R_{\rm{4pt}}(\myeta,\Gamma,\bm b)=\frac{h}{e^2}\frac{\bar{\cal R}(\myeta,\Gamma,\bm b)}{\bar{\cal T}(\myeta,\Gamma,\bm b)},
\end{align}
is an additive self-averaging quantity.  Therefore, in what follows we focus our attention on the four-point resistance. 

It is important to keep in mind that the Landauer theory does contain nonlinear in the applied two-point voltage, $V$, corrections to the conductances and resistances associated with an impurity. However, these corrections are $O(V^2)$. This fact separates them from the nonlinear corrections sought after in this work - magnetochiral anisotropy -  which are $O(b\cdot V)$. 

The mechanism of the appearance of the magnetochiral anisotropy on a helical edge that we consider in this work is associated with the change in the scattering off of a midgap state in the presence of a transport current on the edge. This change is effected by the replacement of the bare midgap energy and the applied magnetic field with their Hartree-Fock-corrected values: 
\begin{align}
    &\myeta\to\myeta+\delta\myeta_{\mathrm{}}\nonumber\\
    &\bm b\to \bm b+\delta\bm b_{\mathrm{}}.
\end{align}
Given that $\delta\myeta_{\mathrm{}}, \,\delta\bm b_{\mathrm{}}$, see Section~\ref{sec:Hfcorrections}, contain $O(V)$ contributions,  their substitution into the formally linear four-point resistance leads to corrections to it that are linear in $V$, if the four-point resistance is expanded in the powers of the voltage. 

Suppressing the dependence of various transport quantities on their arguments for clarity, we obtain the the following $O(V)$ correction to $R_{\rm{4pt}}$ due to the Hartree--Fock corrections: 
\begin{align}\label{eq:4ptRcorrection}
    \delta R_{\rm{4pt}}=\frac{h}{e^2}\frac{1}{\bar{\cal T}^2}
    \left(\partial_\epsilon \bar{\cal R}\, \frac{\partial\delta\myeta_{\mathrm{}}}{\partial V}
    +\partial_{\bm b}\bar{\cal R}\cdot \frac{\partial\delta{\bm b}_{\mathrm{}}}{\partial V}\right)\Big{|}_{V=0}V.
\end{align}
In the above expression we introduced short-hand notations
\begin{align}
    \partial_\epsilon:=\frac{\partial}{\partial \myeta},\,\,\partial_{\bm b}:=\frac{\partial}{\partial \bm b}.
\end{align}
We will constrain ourselves to the regime of small magnetic fields as we believe that the midgap level scattering determines the small-field magnetoconductance. Under this assumption we can set $\bar{\cal T}\to 1$ in the denominator of Eq.~\eqref{eq:4ptRcorrection}. In the same sense we can trade the two-point voltage across the impurity for the transport current flowing on the edge as $V=\frac{h}{e^2}I$. In other words, we used the $\bm b=0$ value of the two-point resistance across a single impurity to write the aforementioned Ohm's law.  This allows us to express the two-point voltage drops across each of the midgap state via the total applied voltage as $V=\frac{h}{e^2}G(0)V_{\rm{tot}}$, where $G(0)$ is the measured value of the two-point conductance on the edge in zero magnetic field. 

The final expression for the $O(V)$ correction to the four-point resistance that we will have to average over disorder realizations reads 
\begin{align}\label{eq:4ptRcorrectionfinal}
    \delta R_{\rm{4pt}}\approx
    \left(\partial_\epsilon \bar{\cal R}\, \frac{\partial \delta\myeta_{\mathrm{}}}{\partial V}
    +\partial_{\bm b}\bar{\cal R}\cdot \frac{\partial{\delta{\bm b}_{\mathrm{}}}}{\partial V}\right)\Big{|}_{V=0}\left[\frac{h}{e^2}\right]^2G(0)V_{\rm{tot}}.
\end{align}

In real quantum spin Hall insulators (QSHIs), such as monolayer $\mathrm{WTe}_2$, there are numerous midgap states along the edge. We will spell out specific assumption about these in Section~\ref{sec:disorderaverage}. Here we would like to obtain the expression for the nonlinear resistance correction for a collection of midgap states, which we will have to average over disorder realization later on. 

Based on the findings of our previous work~\cite{chen2023linearG}, where it was shown that the linear transport is best described by a model of incoherent edge transport, we assume the same regime here. This implies that for a collection of midgap states, labeled with subscript $i$, the total resistance of the edge is given by 
\begin{align}\label{eq:edgeresistance}
    R_{\rm{tot}}=R(0)+\sum_i R_{{\rm{4pt}},i}.
\end{align}
In this equation $R(0)$ is the edge resistance in zero magnetic field. We assume that the smallest part of the edge channel between the source and drain electrodes dominates the measured resistance in the incoherent regime, where the resistance is proportional to the length of the sample.  We do not have anything to say about the origin of the deviation of $R(0)$ from a quantized value for an ideal helical edge, but assume that the mechanism causing it does not change appreciably in magnetic fields of a fraction of Tesla. 

For small magnetic fields, the first term in the right hand side of Eq.~\eqref{eq:edgeresistance} dominates the second one. This means that the impurity averaging can be performed both for the total resistance, and the total conductance, 
\begin{align}\label{eq:edgeconductance}
    G_{\rm{tot}}=\frac{1}{R_{\rm{tot}}}\approx G(0)-G^2(0)\sum_i R_{{\rm{4pt}},i}.
\end{align}
Given the equation for the electric current on the edge, $ I=G_{\rm{tot}}V_{\rm{tot}}$, and using Eqs.~\eqref{eq:4ptRcorrectionfinal}, \eqref{eq:edgeresistance} and~\eqref{eq:edgeconductance}, we see that the nonlinear conductance, defined in Eq.~(1) of the main text, is given by
\begin{align}\label{eq:nonlinearconductance}
    \gamma=-\left[\frac{h}{e^2}\right]^2G^3(0)
    \sum_i\left(\partial_\epsilon \bar{\cal R}_i\, \frac{\partial \delta\myeta_{i}}{\partial V}
    +\partial_{\bm b}\bar{\cal R}_i\cdot \frac{\partial{\delta{\bm b}_{i}}}{\partial V}\right)\Big{|}_{V=0}.
\end{align}
As was discussed in Section~\ref{sec:Hfcorrections}, only the nonequilibrium parts of $\delta\myeta$ and $\delta\bm b$ make a contribution to the part of $\gamma$ that is antisymmetric in the external magnetic field, and defines magnetochiral anisotropy. 

\subsection{Averaging over disorder realizations}\label{sec:disorderaverage}

The expression for the nonlinear conductance~\eqref{eq:nonlinearconductance} pertains to a specific realization of midgap states. Its value varies from sample to sample, since in different samples midgap states have different energies and located at different spatial positions, which translates to different level widths. Below we assume that in the incoherent edge transport regime the additive resistance is a self-averaging quantity, meaning that its average over disorder (midgap states) realization is representative of the the resistance for a given sample. 

Below in this Section we will describe averaging of the edge resistance over spatial positions of the midgap states, as well as their on-site energies. There is another type of disorder likely present in the sample, stemming from the fact that the edge of it is also disordered, and that the direction of the edge spin polarization varies in space. Effectively, this means that the orientation of the external magnetic field with respect to the local $z$-axis in the spin space varies in space. Therefore, disorder averaging should contain averaging over this local $z$-axis direction. We postpone this part of averaging until Section~\ref{sec:angularaverage}, since it is fairly trivial to do.

To define the procedure of averaging over midgap states realizations, we assume that the midgap states are dilute enough so they can be seen as independent scatterers. We take the average number density of the impurities to be constant in space $n(x,y)=n_{\mathrm{mg}}$. We denote the length of the channel as $L_{x}$, and the width of the QSHI sample as $L_{y}$, so on average there are $n_{\mathrm{mg}}L_{x}L_{y}$ impurities with different energies $\myeta$ and different level widths $\Gamma$. Further, we assume that the energies of the midgap states are distributed homogeneously with density $\rho_{\mathrm{mg}}(\myeta)=\rho_{\mathrm{mg}}$. Since the midgap level width due to hybridization with the edge state depends on the distance to the edge, we have to specify this dependence. We take the tunneling matrix element between the edge and a midgap level, and hence its width fall off exponentially with the distance between the impurity and the edge $y$: $\Gamma(y)=\Gamma_{\rm{max}}e^{-\frac{y}{a}}$. Here, $\Gamma_{\rm{max}}$ is the maximum level width that occurs when the mid-gap state is located very close to the edge, and $a$ is a length scale associated with the tunneling between the helical edge states and the midgap level. We note that hybridization with the helical edge state is definitely not the only mechanism that leads to a finite width of the midgap level. So we assume that this hybridization makes the largest contribution to the widths of the levels located in proximity to the physical edge. However, for impurities located far from the edge, their widths must saturate at some minimal value $\Gamma_{\text{min}}$. 

It follows that averaging over midgap states realizations consists of averaging over different numbers of such states, and their parameters for a given number of states. We assume that the distribution of the number of midgap states is Poissonian, with the average given by $n_{\mathrm{mg}} L_x L_y$, and that the average over the parameters of each state are statistically independent. If we denote the average over disorder with $\Left<\ldots\Right>$, we obtain for any quantity $A$ that characterizes an aditive observable
\begin{align}
    \Left<\sum_i A(x_i,y_i,\myeta_i)\Right>=n_{\mathrm{mg}} L_x L_y \int_{0}^{L_{x}} \frac{dx}{L_x} \int_{0}^{L_{y}} \frac{dy}{L_y} \rho_{\mathrm{mg}}\int_{-\infty}^{+\infty} d\myeta\; A(x,y,\myeta).
\end{align}
The first two integrals in the right hand side of the above equation represent the configurational averages over the spatial position of each midgap state, while the integral over $\myeta$ is the average over its on-site energy. We assume translational invariance along the edge, which means that physical observables do not depend of $x$, leading to
\begin{align}
    \int^{L_x}_0\frac{dx}{L_x}\to1.
\end{align}
Within our theory, any dependence of observables on the distance between the midgap state and the physical edge of the system, denoted with $y$, comes from the dependence of the level width on $y$, which for $\Gamma(y)=\Gamma_{\rm{max}}e^{-\frac{y}{a}}$ permits us to formally write
\begin{align}\label{eq:y-integral}
\int^{L_y}_0 \frac{dy}{L_y}=
\frac{a}{L_y}\int^{\Gamma_{\rm{max}}}_{\Gamma_{\rm{max}}e^{-\frac{L_y}{a}}}
\frac{d\Gamma}{\Gamma}.    
\end{align}
In what follows we will adjust the limits of integration in this integral in the following way. We expect $\Gamma_{\rm{max}}$ to be an atomic-scale quantity, pertaining to the situation of a midgap state located very close to the edge. As such, it should be overwhelmingly large as compared to the relevant low energy scales in the problem: $T$, $b$, and so on. Therefore, we can safely set $\Gamma_{\rm{max}}\to \infty$,  given that the integrals we will encounter are convergent in the upper limit, with the integrands exponentially suppressed beyond energies of order $T$. The lower integration limit, $\Gamma_{\rm{max}}e^{-\frac{L_y}{a}}$, is associated with the finite width of the sample, and is unphysically small within our set up. Midgap states located across the sample from the edge of interest should be irrelevant for transport along it. In addition, the width of states located  relatively far from the edge should be determined by mechanisms other than hybridization with the edge. To account for this, we will introduce a phenomenological parameter $\Gamma_{\rm{min}}$ as the lower cut-off of the integral in Eq.~\eqref{eq:y-integral}, which turns it into
\begin{align}
    \int^{L_y}_0 \frac{dy}{L_y}=\to \frac{a}{L_y}\int^{\infty}_{\Gamma_{\rm{min}}}\frac{d\Gamma}{\Gamma}.
\end{align}
This is the form we will use below. 

Given the disorder averaging procedure outlined above, the expression for the disorder-averaged nonlinear conductance, for which we retain the same symbol, becomes
\begin{align}\label{eq:gamma-antisymmetric}
    \gamma_a\approx-\rho_{\rm{mg}}n_{\rm{mg}}L_x a\left[\frac{h}{e^2}\right]^2G^3(0)\int^{\infty}_{-\infty}d\myeta\int^\infty_{\Gamma_{\rm{min}}}\frac{d\Gamma}{\Gamma}
    \left(\partial_\epsilon \bar{\cal R}\, \frac{\partial \delta\myeta_{\rm{neq}}}{\partial V}
    +\partial_{\bm b}\bar{\cal R}\cdot \frac{\partial{\delta{\bm b}_{\rm{neq}}}}{\partial V}\right)\Big{|}_{V=0}.
\end{align}
This expression reduces $\gamma_a$ to an integral. We are only interested in the singular part of this integral for small magnetic fields. Full evaluation of this singularity is cumbersome, hence we will outline the structure of the integral, and give more explicit expressions toward the end of this Section. 

We can interpret the integrand in Eq.~\eqref{eq:gamma-antisymmetric} as a dot product of two vectors in the space spanned by $\myeta,b_x,b_y,b_z$, where the first vector is given by 
\begin{align}
&\left(\partial_\epsilon \bar{\cal R},
    \partial_{\bm b}\bar{\cal R}\right)
    =\int_{-\infty}^{+\infty} dE\; \left[-\frac{\partial n_F\left(E-\mu\right)}{\partial E}\right] \bm F_{1}(E-\myeta, \Gamma, \bm b),
 \end{align}
  and the second one is
    \begin{align}
&\left(\frac{\partial \delta\myeta_{\rm{neq}}}{\partial V}
, \frac{\partial{\delta{\bm b}_{\rm{neq}}}}{\partial V}\right)\Big{|}_{V=0} 
=
-e U
\int_{-\infty}^{+\infty} dE\; \left[-\frac{\partial n_F\left(E-\mu\right)}{\partial E}\right] \bm F_{2}(E-\myeta, \Gamma, \bm b), \label{eq:SecondInt}
\end{align}
where $\bm F_{1,2}$ are some four-dimensional vector functions. We can now write $ \gamma_a=\rho_{\rm{mg}}n_{\rm{mg}}L_x a\left[\frac{h}{e^2}\right]^2G^3(0)eU\gamma_1$, with 
\begin{equation}
    \gamma_{1}= \int_{\Gamma_{\rm{min}}}^{\infty}\frac{d\Gamma}{\Gamma}\int_{-\infty}^{+\infty}d \myeta\int_{-\infty}^{+\infty}d E\int_{-\infty}^{+\infty}d E'
    \left[\frac{\partial n_F\left(E-\mu\right)}{\partial E}\right]
    \left[\frac{\partial n_F\left(E'-\mu\right)}{\partial E'}\right]
    \bm F_{1}(E-\myeta,\Gamma) \cdot \bm F_{2}(E'-\myeta,\Gamma).
\end{equation}
We now proceed with the evaluation of the above integral. We shift $E \to E+\myeta$ and $E' \to E'+\myeta$ and exchange the integration order:
\begin{equation}
    \gamma_{1}=\int_{\Gamma_{\rm{min}}}^{\Gamma_{max}}\frac{d\Gamma}{\Gamma}\int_{-\infty}^{+\infty}d E\int_{-\infty}^{+\infty}d E' \int_{-\infty}^{+\infty}d \myeta
    \left[\frac{\partial n_F(E-\mu+\myeta)}{\partial E}\right]
    \left[\frac{\partial n_F(E'-\mu+\myeta)}{\partial E'}\right]
    \bm F_{1}(E,\Gamma) \cdot \bm F_{2}(E',\Gamma).
\end{equation}

Further, we shift $\myeta \to \myeta-E'+\mu$ to obtain 
\begin{equation}
    \gamma_{1}=\int_{\Gamma_{\rm{min}}}^{\infty}\frac{d\Gamma}{\Gamma}\int_{-\infty}^{+\infty}d E\int_{-\infty}^{+\infty}d E' \int_{-\infty}^{+\infty}d \myeta\left[\frac{\partial n_F(E-E'+\myeta)}{\partial E}\right]
    \left[\frac{\partial n_F(\myeta)}{\partial\myeta}\right]
    \bm F_{1}(E,\Gamma) \cdot \bm F_{2}(E',\Gamma).
\end{equation}
At this point we can evaluate the integral over $\myeta$ using 
\begin{align}
\int_{-\infty}^{+\infty}d z' \: \frac{\partial n_F(z+z')}{\partial z'}\,\frac{\partial n_F(z')}{\partial z'} 
    = \frac{1}{4 T} \text{csch}^2\left(\frac{z}{2 T}\right) \left(-2 + \frac{z}{T} \coth\left(\frac{z}{2 T}\right)\right)=:\frac{1}{T}\Th\left(\frac{z}{T}\right),
\end{align}
and also calculate
\begin{align}
    \bm F_{1}(E,\Gamma) \cdot \bm F_{2}(E',\Gamma)
    =\cos\theta \sin^2\theta \cdot \Sp(E,E',\Gamma, b),
\end{align}
with 
\begin{equation}
\Sp(E,E',\Gamma, b) = \frac{8 b^3 \Gamma ^3 \left(E-E'\right) \left(b^2 \left(E'-E\right)+\Gamma ^2 \left(3E+E'\right)+E^2 \left(E-E'\right)\right)}{\pi \left(\left(b-E\right)^2+\Gamma ^2\right)^2 \left(\left(b+E\right)^2+\Gamma ^2\right)^2 \left(\left(b-E'\right)^2+\Gamma ^2\right) \left(\left(b+E'\right)^2+\Gamma ^2\right)}.
\end{equation}

We proceed by changing variables $\xi=\frac{1}{2}(E+E')$ and $\Xi=E-E'$. Then $\gamma_1$ becomes
 \begin{equation}
    \gamma_{1}={\cos\theta \sin^2\theta}\;\frac{1}{T}\int_{\Gamma_{\rm{min}}}^{\infty}\frac{d\Gamma}{\Gamma} \int_{-\infty}^{+\infty} d \Xi \int_{-\infty}^{+\infty} d\xi \Th\left(\frac{\Xi}{T}\right) \Sp\left(\xi+\frac{\Xi}{2},\xi-\frac{\Xi}{2},\Gamma, b\right).
\end{equation}

In the obtained expression the integral over $\xi$ can be done exactly. However, the result will not be analytic (or even continuous) at $b=0,\Gamma=0$.
So we will perform the integral over $\xi$ explicitly separating out the singular part of the result: 
\begin{equation}
    \int_{-\infty}^{+\infty} d\xi\;\Sp\left(\xi+\frac{\Xi}{2},\xi-\frac{\Xi}{2},\Gamma, b\right) =
    \frac{b^3\Gamma^2}{(b^2+\Gamma^2)^2} \; \Sp_{\rm{I}}(\Xi, \Gamma, b),
\end{equation}
where
\begin{align}
   &\Sp_{\rm{I}}(\Xi, \Gamma, b)= \frac{-2 \Xi ^2 \left(64 \left(b^2+\Gamma ^2\right)
   \left(b^4+10 b^2 \Gamma ^2-7 \Gamma ^4\right)+16 \Xi^2
   \left(3 b^4-18 b^2 \Gamma ^2-5 \Gamma ^4\right) +4 \Xi^4 \left(3 \Gamma ^2-5 b^2\right) +\Xi^6\right)}{\left(4 \Gamma ^2+\Xi ^2\right) \left((\Xi -2 b)^2+4 \Gamma ^2\right)^2
   \left((2 b+\Xi )^2+4 \Gamma ^2\right)^2}\label{suppeq:SpI}.
\end{align}
Now $\gamma_{1}$ can be written as
\begin{align}
    \gamma_{1}=&{\cos\theta \sin^2\theta }\;\frac{1}{T}\int_{\Gamma_{\rm{min}}}^{\infty}d\Gamma\;\frac{b^3\Gamma}{(b^2+\Gamma^2)^2} \int_{-\infty}^{\infty}d \Xi\; \Th\left(\frac{\Xi}{T}\right)\; \Sp_{\rm{I}}(\Xi,\Gamma,b) \nonumber
    \\
    =&{\cos\theta \sin^2\theta }\;\frac{1}{T^2}\int_{\Gamma_{\rm{min}}}^{\infty}d\Gamma\;\frac{b^3\Gamma}{(b^2+\Gamma^2)^2} \int_{-\infty}^{\infty}dz\; \Th\left(z\right)\; \Sp_{\rm{I}}\left(z,\frac{\Gamma}{T},\frac{b}{T}\right) \label{eq:gamma1integral}
\end{align}
by changing variables $z = \Xi/T$ and using the fact that $\Sp_{\rm{I}}$ is a homogeneous rational function of degree $-2$ (so essentially dimensional analysis).

We further define the result of the $z$ integration as a function $F$ defined on $(0,\infty) \times (0,\infty)$ via
\begin{align}
 \forall \Tilde{\Gamma}>0, \Tilde{b}>0, \quad
    F(\Tilde{\Gamma},\Tilde{b}) :=& \int_{-\infty}^{\infty}d z\; \Th(z)\; \Sp_{\rm{I}}(\Xi,\Tilde{\Gamma},\Tilde{b}) \label{eq:Fdefinition}
\end{align}
Note $F$ is to be thought of as a function of dimensionless variables, and eventually $\Tilde{b}$ will be replaced by $b/T$ etc.
We then claim (without proof, although it appears so numerically) that this can be continuously extended to an analytic function on $[0,\infty) \times [0,\infty)$, and will compute an exact result for $F(0,0)$.
Note that if we had \textit{not} separated off the singular part, we would not be able to do this, since $\lim_{(b,\Gamma) \to (0,0)} \frac{b^3\Gamma}{(b^2+\Gamma^2)^2}$ does not exist.

The fact that $F$ can be continuously extended is non-trivial. Note that Eq.~\eqref{eq:Fdefinition} \textit{cannot} be used to define $F(0,0)$ directly, since $\Sp_{\rm{I}}(z,0,0)=-2/z^2$ and so $\Th(z)\Sp_{\rm{I}}(z,0,0)$ is \textit{not} integrable, due to the singularity at $z=0$. The trick to compute $F(0,0))$ is to rewrite the integral in way that we \textit{can} simply exchange the limit with the integral. Note that pictorially, you can visualize $\Sp_{\rm{I}}(z,\Tilde{\Gamma},\Tilde{b})$ as having `structure' at small $z\sim \Tilde{\Gamma},\Tilde{b}$. We then note that the thermal kernel $\Th(z)$ changes on the scale of $z\sim 1$ and that $\Th(0)=1/6$.
So for small (but nonzero) $\Tilde{b}$, $\Tilde{\Gamma}$,
we might to first approximation set $\Th(z) \to \Th(0)$.
The resulting rational function can be integrated exactly, and it can be shown that it ends up simply yielding 0.
\begin{align}
 \forall \Tilde{\Gamma}>0, \Tilde{b}>0, \quad
    \int^{\infty}_{-\infty} dz\; \Sp_{\rm{I}}(z,\Tilde{\Gamma},\Tilde{b})=0
\end{align}
Thus we, without changing the value of the integral, rewrite our integrand
\begin{align}
 \forall \Tilde{\Gamma}>0, \Tilde{b}>0, \quad
    F(\Tilde{\Gamma},\Tilde{b}) &=
    \int^{\infty}_{-\infty} d z\; \left(\Th(z)-\Th(0)\right)
    \Sp_{\rm{I}}(z,\Tilde{\Gamma},\Tilde{b})
\end{align}
This is good, because now setting $\Tilde{\Gamma}$,$\Tilde{b}$ to zero in the integrand \textit{does} yield something integrable.
Thus we compute:
\begin{align}
    F(0,0) =&
    \int^{\infty}_{-\infty} d z\; \left(\Th(z)-\Th(0)\right) \Sp_{\rm{I}}(z,0,0) \\
    =& 2 \int^{\infty}_{-\infty} d z\; \frac{\left(\Th(0)-\Th(z)\right)}{z^2}\\
    =& \frac{3\zeta(3)}{\pi^2},
\end{align}
where $\zeta(3)$ is the Riemann-Zeta-function at $3$.

Going back to \eqref{eq:gamma1integral}, we obtain \footnote{
Note that the limit $b \to 0, \Gamma_{min} \to 0$ depends what order you take it. And since we expect there to be some physical $\Gamma_{min}$, there will be regimes of applied $b$ which are both above and below it.
In the $b\ll\Gamma_{min}$ regime the lowest order dependence is $b^3$,
while for $\Gamma_{min}\ll b$ the dependence is linear in $b$, for still $b\ll T$.
}
\begin{align}
    \gamma_1 \approx \frac{3\zeta(3)}{\pi^2}\cos\theta \sin^2\theta\;\frac{1}{T^2} \int_{\Gamma_{\rm{min}}}^{\infty}d\Gamma \frac{b^3\Gamma}{(b^2+\Gamma^2)^2}  =  \frac{3\zeta(3)}{2\pi^2}\cos\theta \sin^2\theta\;\frac{1}{T^2}\frac{b^3}{\Gamma_{\rm{min}}^2+b^2},
\end{align}
where the approximation here is in replacing $F\left(\frac{\Gamma}{T},\frac{b}{T}\right)$ with $F(0,0)$ inside the $\Gamma$ integral. This approximation is justified since a Taylor expansion of $F$ will lead to terms which, when properly regularized, will be higher order in $\frac{1}{T}$.\footnote{
Note that $F\left(\frac{\Gamma}{T},\frac{b}{T}\right)$
will begin to cut off exponentially around
$\Gamma\sim T$. Taylor expanding $F$ will lead to divergent integrals for the individual terms in the expansion. But we can see what kind of contribution they could make to the integral by replacing the $\infty$ upper limit with a cutoff $\Lambda\sim T$.
Doing so we see in both regimes only corrections higher order in $\frac{b}{T}$ and $\frac{\Gamma_{\rm{min}}}{T}$
}

Going back to $ \gamma_a=\rho_{\rm{mg}}n_{\rm{mg}}L_x a\left[\frac{h}{e^2}\right]^2G^3(0)eU\gamma_1$,  we get
\begin{equation}\label{eq:gamma-antisymmetric-final}
    \gamma_a(\bm{b})
 =\frac{3\zeta(3)}{2\pi^2}\frac{h^{2}}{e^{4}} \rho_{\rm{mg}}n_{\rm{mg}} L_{x} a  G^{3}(0)
  \frac{eU}{T^2}
   \frac{b^3}{\Gamma_{\rm{min}}^2+b^2}
\cos\theta \sin^2\theta.
\end{equation}
Using this result for $\gamma_a(\bm b)$, one can recover various results used in the main text.

\subsection{Angular distribution due to uneven monolayer $\rm{WTe}_{2}$ }\label{sec:angularaverage}
In this section, we take into account phenomenologically randomness in the orientation of the helical edge spin polarization, keeping in mind the  monolayer $\rm{WTe}_{2}$ as the system of interest. In that case the edge spin polarization lies in the mirror plane of the sample, and makes angle $\theta_0$ with the $z$-axis, defined as the normal to the sample's plane. For constant $\theta_0$, \textit{i.e.} independent of $x$, one can simply shift $\theta\to \theta-\theta_0$ in Eq.~\eqref{eq:gamma-antisymmetric-final} to describe the magnetochiral anisotropy. 

We model disorder in the spin polarization by making $\theta_0$ coordinate-dependent,  $\theta_0\to\theta_0(x)$, but assume that the spin polarization stays in the mirror plane of the sample. This is the simplest model, not necessarily realistic, which still illustrates the effect of such type of disorder.  We further assume that $\theta_0(x)$ does not change much in the scattering region near each midgap state, but is uncorrelated between different scattering regions. This situation correspond to each midgap state's contribution to the resistant depending on the local orientation of the spin polarization, which can be considered as another midgap state parameter one needs to average over. 

We suppose that the distribution of the local azimuthal angles is normal, with mean value $\bar\theta_{0}$ and standard derivation $\sigma$: 
\begin{equation}
    P(\theta_0)=\frac{1}{\sqrt{2 \pi}\sigma}e^{-\frac{(\theta_0-\bar\theta_{0})^{2}}{2 \sigma^2}}.
\end{equation}
Disorder averaging then acquires another step, which amounts to averaging over $\theta_0$. This alters the angular dependence of the magnetochiral anisotropy, see Eq.~\eqref{eq:gamma-antisymmetric-final}: 
\begin{align}
   \cos (\theta-\theta_0) \sin^{2} (\theta-\theta_0)\to \int_{0}^{\pi}d \theta_0 P(\theta_0) \cos (\theta-\theta_0) \sin^{2} (\theta-\theta_0) .
\end{align}
We now introduce  $\phi=\theta_0-\bar\theta_0$, and assume that $\sigma\ll$1, such that the integral over $\phi$
can be extended to $(-\infty,\infty)$. As a result, we obtain 
\begin{align}
    \int_{0}^{\pi}d \theta_0 P(\theta_0) \cos (\theta-\theta_0) \sin^{2} (\theta-\theta_0)\approx\int_{-\infty}^{\infty}d \phi \frac{1}{\sqrt{2 \pi}\sigma}e^{-\frac{\phi^{2}}{2 \sigma^2}} \cos (\theta-\bar\theta_0-\phi) \sin^{2} (\theta-\bar\theta_0-\phi).
\end{align}
Performing the integral is elementary, and for $\sigma\ll1$ results in 
\begin{equation}
    \gamma_a(\bm b) \approx  \frac{3\zeta(3)}{2\pi^2}\frac{h^{2}}{e^{4}} \rho_{\rm{mg}}n_{\rm{mg}} L_{x} a  G^{3}(0)
  \frac{eU}{T^2}
    \frac{b^3}{\Gamma_{\rm{min}}^2+b^2}
    \, \cos (\theta-\bar\theta_0) \left[\sin^{2} (\theta-\bar\theta_0)+\delta\right],
\end{equation}
where $\delta \approx \sigma^2$, and we neglected a small renormalization of the overall magnitude of the effect. This is the form of the magnetochiral anisotropy used in Eq.~(8) of the main text. 

\section{Band mechanism of magnetochiral anisotropy} 

There is another mechanism of MCA on a topological edge, which is related to the nonlinearity of the edge dispersion. It is similar in spirit to the analogous effects in carbon nanotubes~\cite{spivak2002nanotube}, and topological insulator surface states~\cite{vignale2018bilinear}.  The detailed theory of this effect will be reported elsewhere~\cite{chenunpublished}. Here we present a brief symmetry-based argument for it, and show that taking it into account amounts to redefinition of constant $\delta$ in Eq.~\eqref{eq:gammafinal}. 

The general single-particle Hamiltonian for a translationally-invariant helical edge state with energy-independent spin polarization direction $\bm d_{so}$ reads
\begin{align}
    h_{k}=\epsilon_s(k)+\epsilon_a(k)\bm \sigma\cdot \bm d_{so},
\end{align}
where $\epsilon_{s,a}(-k)=\pm \epsilon_{s,a}(k)$ is required by time-reversal symmetry. In the presence of a Zeeman field $\bm b=b_\parallel \bm d_{so}+\bm b_\perp$, with obvious notations for the components parallel and perpendicular to $\bm d_{so}$, the Hamiltonian becomes 
\begin{align}
    h_{k}=\epsilon_s(k)+(\epsilon_a(k)+b_{\parallel})\bm \sigma\cdot \bm d_{so}+\bm b_\perp\cdot\bm \sigma,
\end{align}
which has conduction and valence eigenbands of energies $\epsilon_{c,v}=\epsilon_{s}(k)\pm\sqrt{(\epsilon_a+b_\parallel)^2+b_\perp^2}$. It is clear that for $b_\parallel\neq 0$, the band dispersions are asymmetric in the edge Brillouin zone: $\epsilon_{c,v}(k)\neq \epsilon_{c,v}(-k)$, \textit{i.e.} the spectrum lacks an inversion center in the edge Brillouin zone. This is a general situation for broken inversion and time-reversal symmetries. 

To the linear order in the accelerating electric, there is redistribution of the charge carriers between the left- and right-moving branches of the band crossing the chemical potential. To the quadratic order in the electric field there is also redistribution of the carriers within each branch, without changing the total number on each branch. If the branch dispersion is non-linear, this redistribution changes the net current carried by the branch, and if the inversion symmetry is broken in the momentum space, these current changes do not cancel between the two branches, leading to a current change quadratic in the electric field, and linear in $b_\parallel$, which breaks the time reversal, and the inversion in the momentum space. 

This argument shows that this band mechanism leads to a contribution to the magnetochiral anisotropy $\gamma(\bm b)\propto b_\parallel\propto \cos\theta$ for weak magnetic fields. This contribution modifies $\sin^2\theta\to \sin^2\theta+\rm{const}$ in Eq.~\eqref{eq:gamma-antisymmetric-final}, but alone is insufficient to explain the full angular dependence of the magnetochiral anisotropy.

\clearpage
\bibliography{Citation.bib}
\bibliographystyle{apsrev4-2}

\appendix